%% file: main.tex
\documentclass[a4paper, preprint]{elsarticle} 
\input{sections/0-preambule}

\title{\corr{Fast reconstruction of atomic-scale STEM-EELS images from sparse sampling}\tnoteref{t1}}

\author[1]{Etienne Monier\corref{cor1}}
\ead{etienne.monier@enseeiht.fr}

\author[1,2]{Thomas Oberlin}
\ead{thomas.oberlin@enseeiht.fr}

\author[3]{Nathalie Brun}
\ead{nathalie.brun@u-psud.fr}

\author[3]{Xiaoyan Li}
\ead{xiaoyan.li@u-psud.fr}

\author[3]{Marcel Tenc\'e}
\ead{marcel.tence@u-psud.fr}

\author[1,4]{Nicolas Dobigeon}
\ead{nicolas.dobigeon@enseeiht.fr}

\address[1]{University of Toulouse, IRIT/INP-ENSEEIHT, 31071 Toulouse Cedex 7, France}
\address[2]{University of Toulouse, ISAE-SUPAERO, 31400 Toulouse, France}
\address[3]{Université Paris-Saclay, CNRS, Laboratoire de Physique des Solides, 91405, Orsay, France}
\address[4]{Institut Universitaire de France (IUF), France}
\cortext[cor1]{Corresponding author}

\tnotetext[t1]{Part of this work was presented at the SPARS 2019 workshop, Toulouse, France \cite{MonierSPARS2019}. Part of this work has been funded by CNRS, France, through the Imag’in ARSIS project, the METSA 17A304 program and the ANR-3IA Artificial and Natural Intelligence Toulouse Institute (ANITI) under grant agreement ANR-19-PI3A-0004. This project has received funding from the European Union's Horizon 2020 Research and Innovation Programme under grant agreement no 823717.}

\begin{document}

\begin{abstract}
This paper discusses the reconstruction of partially sampled spectrum-images to accelerate the acquisition in scanning transmission electron microscopy (STEM). 
The problem of image reconstruction has been widely considered in the literature for many imaging modalities, but only a few attempts handled 3D data such as spectral images acquired by STEM electron energy loss spectroscopy (EELS).
Besides, among the methods proposed in the microscopy literature, some are fast but inaccurate while others provide accurate reconstruction but at the price of a high computation burden. Thus none of the proposed reconstruction methods fulfills our expectations in terms of accuracy and computation complexity. 
In this paper, we propose a fast and accurate reconstruction method suited for atomic-scale EELS. 
%
This method is compared to popular solutions such as beta process factor analysis (BPFA)  which \corr{is used for the first time on STEM-EELS images}. %
%
Experiments based on real as synthetic data will be conducted.
\end{abstract}


\maketitle

\input{sections/1-introduction}

\input{sections/2-related-works}

\input{sections/3-proposed-method}


\input{sections/4-experiments}

\input{sections/5-results}

\section{Conclusion}\label{sec:conclusion}

In this paper, we introduced a new reconstruction method referred to as \CLS{} for STEM-EELS imaging that is both fast and accurate. Experiments on synthetic and real data showed it was much faster than dictionary learning-based reconstruction methods and \corr{more accurate} than NN methods. The combination of these two advantages is highly interesting to envisage its practical implementation into an online experimental setup. 

Moreover, time-consuming but more accurate dictionary learning-based methods were also used as a post-processing of the results provided by \CLS{}. These methods performed reconstruction as a 3D task and gave excellent reconstruction quality. However, the price to pay was a high computational time, which could be reduced by using the reconstructed spectrum-image recovered by \CLS{} as an initialization.

Conducting online microscopy acquisition and reconstruction is an active research task as it speeds up acquisition procedures and identification of components. Performing reconstruction with \CLS{} as an online algorithm and coupling it with adaptive sampling is an interesting perspective towards dynamic STEM-EELS imaging. \me{Such an acquisition protocol would allow sample drifts to be handled, which was not the case in the experiments reported in this paper.}

\section*{Acknowledgments}

\corr{Authors would like to thank Dr. Daniele Preziosi, University of Strasbourg, France, for the thin film synthesis and Dr. Alexandre Gloter, Universit\'e Paris-Saclay, France, for the FIB lamella preparation. They are also grateful to the anonymous reviewers for relevant comments which help to improve this manuscript.}

\begin{appendices}
\input{sections/A-FS-implementation}

\input{sections/B-Synthetic-details}

\end{appendices}

\bibliographystyle{elsarticle-num} 
\bibliography{bib/strings_all_ref,bib/bib}

\end{document}

%% file: sections/0-preambule.tex
\usepackage{styles/Notations}

\usepackage[utf8]{inputenc}

\usepackage{amsmath}            
\usepackage{amssymb}            
\usepackage{amsfonts}           
\usepackage{mathrsfs}           
\usepackage{dsfont}             

\usepackage{graphicx}
\usepackage{subfigure}

\usepackage{tikz,pgfplots}
\pgfplotsset{compat=1.13}
\usepackage{pgfplotstable}

\usepackage[ruled,vlined,linesnumbered]{algorithm2e}

\usepackage{algpseudocode}

\usepackage{tabularx}
\usepackage{array}
\usepackage{multirow}
\usepackage{booktabs}  

\usepackage{xcolor}
\usepackage{colortbl}  
\usepackage{styles/colorbrewer}

\usepackage{calc}
\usepackage{forloop}
\usepackage{python}
\usepackage{ifthen}

\usepackage{comment}
\usepackage[page]{appendix}
\usepackage[hidelinks]{hyperref} 
\hypersetup{pdfauthor=author}  



\usepackage{fontawesome}

\newcommand{\minusfa}[1][1]{%
    \forLoop{1}{#1}{ctfa}{%
        {\color{red}\faicon{minus-square}}
    }%
}
\newcommand{\plusfa}[1][1]{%
    \forLoop{1}{#1}{ctfa}{%
        {\color{green}\faicon{plus-square}}
    }%
}

\usepackage{styles/Notations}

\usetikzlibrary{backgrounds}
\usepgfplotslibrary{fillbetween}
\usepgfplotslibrary{colorbrewer}
\usepgfplotslibrary{groupplots}
\usetikzlibrary{matrix}
\usetikzlibrary{spy}
\usetikzlibrary{shadows}

\pgfplotsset{cycle list/Dark2}

\pgfplotsset{every axis/.append style={%
    axis line style = {semithick},
    axis x line = bottom,
    axis y line = left,
    grid=both,
    grid style={line width=.1pt, draw=gray!10},
    major grid style={line width=.2pt,draw=gray!50},
    ticklabel style={font=\tiny},  
    enlargelimits=true,
    legend style={draw=none},
    legend cell align={left},
}}%

\pgfplotsset{every axis plot/.append style={line width=1pt,}}

\usepackage{tikzscale} 
\usepackage{lipsum}

\newlength{\mysize}

\newlength\mycolwidth

\newlength\myfigurewidth

\def\externalizedflag{false}

\newcommand{\extcode}[2]{%
  \ifthenelse{\equal{\string \externalizedflag}{\string true}}
    {#1}
    {#2}%
}

\extcode{%
    \newcommand{\externaldirectory}{external/}
    \usepgfplotslibrary{external} 
    \tikzexternalize[prefix=\externaldirectory]
}{}

\newcommand{\corr}[1]{\textcolor{black}{#1}}

\newcommand{\me}[1]{\textcolor{black}{{#1}}}

%% file: sections/1-introduction.tex

\section{Introduction}

Electron energy loss spectroscopy (EELS) performed in a scanning transmission electron microscope (STEM) has proved to be a powerful tool to analyze chemical components and structures of a sample with a sub-nanometer spatial resolution.  A focused electron probe is scanned over the sample and for each probe position an EELS spectrum  is acquired, as well as several other signals \corr{such} as high-angle annular dark field (HAADF).  The spectrum-image thus acquired can be used to build not only maps of the spatial distribution of the elements but also maps of edges' fine structures corresponding to local electronic structures.

These important capabilities of modern microscopes are somewhat limited by sample damage, instabilities and poor signal-to-noise ratio (SNR). Indeed, acquiring such EELS data set requires a suitable SNR and typical EELS dwell time (exposure time per location) are in the ms range ($1-100$ms). These long dwell times proportionally lead to a significant total \corr{electron dose} received by the sample. This dose increases potential radiation damages \corr{to} the sample~\cite{egerton2004radiation}. This is particularly problematic for sensitive materials such as biological samples.    
Moreover a long acquisition time may increase image distortions caused by time-dependant instabilities of the sample and the microscope. In particular, these instabilities may be substantial at atomic scales. \corr{Performing a multi-frame acquisition, followed by a non-rigid alignment step~\cite{zobelli2019spatial}, is a promising research domain to improve the spatial resolution and to reduce beam-induced damage. The new generation of direct detection cameras with negligible correlated noise could promote the use of this multi-frame setup with even lower dwell-times}.
Finally some high resolution acquisitions need to cover large areas (such as in \cite{anderson2013sparse} for \corr{scanning electron microscopy} (SEM)), leading to long acquisition total time, heavy data storage and long processing steps. 
To increase acquisition speed and/or reduce the full beam exposure, a solution consists in reducing dwell time and subsequently denoising the data as a post-processing operation. Yet, reducing the exposure may be of limited interest since the resulting SNR becomes too low to expect good denoising performance, especially in the case of fine structure analysis.

A recent popular alternative  is \corr{\emph{sparse (or partial) sampling}}. This strategy consists in acquiring the relevant signals only in a small proportion of spatial locations, \corr{which allows for higher dwell time at these positions resulting in the same amount of total electron dose}. The resulting acquired image is partially empty and a reconstruction step is required to obtain a fully exploitable image.
%
This paradigm received a renewed interest since the theoretical results of \emph{compressed sensing} (CS) which states that exact recovery of sub-Nyquist rate acquisitions is possible under certain conditions -- one of them is that the data should be sparse in an appropriate basis. The CS paradigm states that the data should be projected on $n$ random subspaces with $n$ far below the data size, which is well adapted to electron microscopy tomography~\cite{binev2012compressed, jacob2019MM, jacob2018MM}. 
These results raised a lot of interest toward inverse problems which estimate the image based on partial spatial acquisitions which is referred to as \emph{inpainting}. It remains an active research area for STEM~\cite{beche2016compressed,stevens2013potential} and \corr{SEM}~\cite{anderson2013sparse}, among others.

The two previously described acquisition schemes have pros and cons. Schematically, low dwell-time acquisition usually produces better spatial results while \corr{sparse sampling} images usually have rich spectral information. Determining which approach is the best is not trivial. 
%
%
To that end, recent works studied and compared these solutions~\cite{trampert2018ultramicroscopy, monier2018tci} based  on experiments conducted on synthetic as well as real images.

Following the second aforementioned acquisition scheme, this paper addresses the problem of reconstructing spatially sub-sampled atomic-scale STEM EELS images. In particular one motivation here aims at reducing computational burden of the inpainting procedure to make its future implementation possible into the acquisition process. The experimenter should be able to visualize the full spectrum-image along the acquisition, which requires both fast computation and a good accuracy. In addition to this online setup, the experimenter should be able to refine the reconstructed spectrum-image afterwards, where very accurate but possibly time-consuming algorithms are allowed. 
%
\corr{To that end, we propose a new reconstruction method exhibiting a relevant trade-off between accuracy and complexity. We will also show that this proposed technique can serve as a good initialization to accelerate more efficient yet more computationally intensive methods. Moreover, among the compared methods, we propose to apply the popular beta process factor analysis (BPFA), originally dedicated to remote sensing images~\cite{xing2012siam}. Up to our knowledge, it is the first time BPFA is applied to STEM-EELS images, although it was already used in many microscopy works for 2D data restoration such as in SEM~\cite{trampert2018ultramicroscopy}.}
%
The paper is organized as follows. Section~\ref{sec:related-works}  presents an overview of inpainting techniques already used in electron microscopy, with the emphasis \corr{placed} on 2D and 3D reconstructions. Since no fast and accurate 3D method fulfills all requirements to envisage a fully operational online implementation, Section~\ref{sec:proposed-method} describes the newly proposed reconstruction method. \corr{Section~\ref{sec:experiments} describes the synthetic, semi-real and real data, as well as the  experiments conducted to compare the proposed approach to previous works (especially BPFA as a 3D reconstruction method). The experimental results are reported and discussed in Section~\ref{sec:results}}. Section~\ref{sec:conclusion} concludes this study.

%% file: sections/2-related-works.tex

\section{Related works}\label{sec:related-works}

The focus of this paper deals with reconstructing spatially sub-sampled STEM images. Many works considered this problem with different methods and modalities. Most of them were proposed to process single 2D images while few considered the reconstruction of 3D images. This section discusses these works which are mainly divided into two parts. The first one considers \emph{learning-free} methods which reconstruct the images based on the single acquired dataset. The second one studies \emph{learning-based} methods which capitalize on a learning set to calibrate an operator subsequently used to reconstruct the data.

\subsection{Learning-free methods}\label{subsec:learning-free-methods}

Among learning-free methods, nearest neighbor (NN) interpolation is a fast and simple solution possibly allowing for dynamic joint acquisition and reconstruction. To avoid piecewise-constant image like reconstruction, preferred solutions interpolate with a weighted mean over a neighborhood. The weights are chosen to be the normalized inverse distance between the interpolated pixel and the neighbor for reconstructing SEM data in~\cite{godaliyadda2018tci}, energy-dispersive X-ray spectroscopy (EDS) data in~\cite{zhang2018reduced,hujsak2018high} and EELS data in~\cite{hujsak2018high}. An alternative considered for SEM images in~\cite{trampert2018ultramicroscopy} was based on a natural neighbor interpolation, which adjusts the weights based on a Voronoï cell representation \cite{sibson1981natural}. Note that STEM acquires full spectra at particular spatial positions.

Regularized least-square (LS) methods generally offers better results than NN as they additionally constrain the reconstructed image to fulfill a predefined behavior, usually promoted by a well chosen regularization.
Akin to the CS paradigm, a classical regularization is the sparsity of the reconstructed image in an appropriate basis, as the $\ell_1$-norm regularized LS problem considered in~\cite{han2018optimal} for \corr{atomic force microscopy (AFM)}. This type of regularized LS m\corr{e}thods will be referred as $\ell_1$-LS in Table~\ref{table:available-methods}.
In the case of periodic structures (as for atomic-scale images), this basis can be  Fourier or  discrete cosine transformation (DCT). The authors in~\cite{stevens2018apl} proposed an inpainting method for atomic-scale \corr{high-angle annular dark-field (HAADF)} images based on a thresholded Fourier transform, which constrains the image sparsity in periodic basis. 
The method in~\cite{Beche2016development} promoted the sparsity of the DCT representation to reconstruct \corr{HAADF} images, using the SPGL1 algorithm~\cite{berg2008probing}. In the same way,  this regularization can be coupled with a wavelet basis to dynamically reconstruct HAADF data \cite{li2018compressed}.
Another standard regularization is total variation (TV), i.e., the $\ell_1$-norm of the image gradient promoting piecewise constant reconstructed image, as considered in~\cite{han2018optimal} for AFM. The block-DCT representation was coupled with TV for reconstructing SEM data in \cite{anderson2013sparse}. The $\ell_2$-norm of the image gradient is also widely used as a regularization to promote spatial smoothness and is referred to as the Sobolev energy \cite{monier2018tci}.
In the case of multi-band images, spectral regularizations were proposed in addition to the spatial one. In~\cite{monier2018tci}, for instance, the 3S method uses the weighted $\ell_2$-norm across the EELS spectrum-image bands or simply a nuclear norm (which ensures the low-rank nature of the reconstructed data) in addition to the classical Sobolev energy spatial regularization.

Another class of reconstruction methods exploit the spatial redundancy in the image, often referred to as \emph{patch-based methods}.
They form a very popular and successful class of reconstruction methods which raised a lot of attention in the last decades to solve  inverse problems such as denoising, inpainting and deblurring. 
%
For example, the exemplar-based inpainting~\cite{criminisi2004region} (EBI) \corr{reconstructs partially corrupted images by iteratively replacing the image patches by the best matching uncorrupted patch extracted from its neighborhood.}
%
To describe spatial redundancy, successful algorithms aims at sparsely representing image patches thanks to atoms of a dictionary jointly learned with the reconstructed data. BPFA is probably the most popular dictionary learning (DL) method in the microscopy community~\cite{xing2012siam}. It was first used for atomic-scale HAADF images in~\cite{stevens2013potential} and was used afterwards in many papers for the same kind of data~\cite{mucke2016practical,kovarik2016implementation}.
The authors of BPFA proposed Kruskal-factor analysis (KFA) as a tensor extension of BPFA~\cite{stevens2017tensor}. KFA was used to reconstruct EELS images based on a multiplexed spectrum-image acquisition~\cite{stevens2016mm}.
Last, the expected-patch log-likelihood (EPLL) algorithm assumes the patch distribution to be a Gaussian mixture ~\cite{zoran2011from}. Yet, its computation time is important and the authors in~\cite{hujsak2018high} preferred a simplified but accelerated version called fast EPLL (FEPLL) to reconstruct SEM images \cite{parameswaran2019accelerating}.
In addition to the patch-based methods used in the microscopy community, wKSVD~\cite{mairal2008tip} and ITKrMM~\cite{naumova2018fast,naumova2017dictionary} learn the dictionary from incomplete data without assuming particular patch distribution. They remain state-of-the-art methods that will be considered in this paper.

To achieve better performance with reduced acquisition time, several predefined scan patterns were proposed such as regular scan~\cite{monier2017reconstruction}, random horizontal lines~\cite{kovarik2016implementation,han2018optimal}, mixed regular-random scan~\cite{stevens2018apl,monier2017reconstruction}, spiral scans~\cite{sang2017dynamic,li2018compressive,han2018optimal} or square-shape scan~\cite{han2018optimal}.
%
These results tend to show that the best \corr{performance is} achieved by semi-random scan patterns, which introduce randomness and avoid large holes. %
\corr{Last, adaptive sparse scanning enables consequent reconstruction improvement by selecting the pixel to sample based on previously acquired data. In~\cite{dahmen2016feature}, the authors proposed to perform a first low-SNR scan to locate the spatial edges. A second high-SNR scan is then performed on these edges only. Finally, in the low-SNR acquired image, the smooth regions are filtered and the edges are filled with the pixels from the second high-SNR acquisition. An alternative adaptive scanning scheme proposed in \cite{dahmen2019adaptive} consists in iteratively locating possible points of interest to be sampled. Learning-based adaptive sparse scanning are discussed in the next subsection}.

\subsection{Learning-based methods}

Contrary to learning-free methods which recover the full image based only on the partially acquired data, learning-based methods learn an operator based on a possibly large training set. 
\corr{These methods are known to be much more accurate as long as the geometric content of the image to reconstruct is similar to the content of the training data.}

For instance, the GOAL algorithm learns a dictionary which maximizes the sparsity of the training dataset representation~\cite{hawe2013analysis}. The learned dictionary is then used to perform the inpainting task for test data. \corr{Similarly, EBI which is originally a learning-free method, can be adapted to benefit from the availability of a training set. To that end, within the conventional EBI framework, instead of extracting the copied patch from the neighborhood, this patch can be chosen from  a dictionary learned beforehand on uncorrupted images. This is the strategy followed in~\cite{trampert2018exemplar} to reconstruct 3D SEM data. GOAL and the learning-based counterpart of EBI were used} in~\cite{trampert2018ultramicroscopy} for 2D SEM images but BPFA seemed to give better results. 

Learning-based approaches can also be designed to decide which positions should be sampled in order to minimize the distortion after reconstruction. Indeed, sub-sampled data reconstruction performance highly depends on the sample locations \cite{trampert2018ultramicroscopy}. 
To improve the reconstruction quality, the supervised learning approach for dynamic sampling (SLADS) learns a function (called \emph{expected reduction in distortion} (ERD)) indicating which location should be sampled to maximally reduce distortion \cite{godaliyadda2018tci}.
This learning step is based on a list of descriptors and requires labeled training data, and it was used to dynamically sample SEM images.
This method has been also applied to EDS data in~\cite{zhang2018reduced}. To that end, a convolutional neuronal network (CNN) classifies the test data spectra and the ERD function is computed simultaneously for all labels.
The paper ~\cite{hujsak2018high} modified this approach to allow mixed elements in EELS and EDS.
All these approaches needs a fast reconstruction which is achieved thanks to a weighted NN technique.

\begin{table*}
    \centering
    \resizebox{\textwidth}{!}{ 
    \bgroup
        \renewcommand{\arraystretch}{1.2}
        \begin{tabular}{>{\arraybackslash\centering}m{3cm}*{5}{c}}
            \toprule
            \multirow{2}*{Family}&
            \multirow{2}*{Method}&
            \multicolumn{2}{c}{Works}&
            \multirow{2}*{Execution time}&
            \multirow{2}*{Accuracy}\\
            &&2D&3D&&\\
            \midrule
            \multirow{2}*{NN}&
            NN&
            -&
            -&
            \plusfa[3]&
            \minusfa[2]\\
            &
            Weighted neighbor&
            \cite{sibson1981natural,trampert2018ultramicroscopy}&
            -&
            \plusfa[2]&
            \minusfa[1]\\
            \midrule
            \multirow{4}*{LS-regularized}&
            $\ell_1$-LS& 
            \cite{han2018optimal,Beche2016development,li2018compressed,anderson2013sparse}&
            &
            \plusfa&
            \plusfa\\
            &
            TV-LS& 
            \cite{han2018optimal}&
            -&
            \plusfa[1]&
            \plusfa\\
            %
            %
            &
            3S&
            -&
            \cite{monier2018tci}&
            \plusfa&
            \plusfa\\
            \midrule
            \multirow{6}{3cm}{\centering DL-based methods}&
            BPFA&
            \corr{\cite{stevens2013potential,trampert2018ultramicroscopy}}&
            \textit{\cite{xing2012siam}}&
            \minusfa[3]&
            \plusfa[3]\\
            &
            EBI&\cite{trampert2018ultramicroscopy}&
            \corr{\cite{trampert2018exemplar}}&
            \minusfa[1]&
            \plusfa[2]\\
            &
            FEPLL&
            \textit{\cite{parameswaran2019accelerating}},\cite{hujsak2018high}&
            -&
            \minusfa[1]&
            \plusfa[2]\\
            &
            wKSVD&
            -&\textit{\cite{mairal2008tip}}&
            \minusfa[2]&
            \plusfa[2]\\
            &
            ITKrMM&
            \textit{\cite{naumova2018fast}}&
            \textit{\cite{naumova2017dictionary}}&
            \minusfa[1]&
            \plusfa[2]\\
            &
            GOAL&
            \textit{\cite{hawe2013analysis}},\cite{trampert2018ultramicroscopy}&
            -&
            \minusfa[1]&
            \plusfa[2]\\
            \bottomrule
        \end{tabular}
    \egroup
    }
    \caption{Comparison of the methods proposed in the microscopy literature for reconstructing partially sampled images. \corr{Additional references not originated from the microscopy literature are also provided in italics.} The execution time and accuracy are qualitatively evaluated based on the results of Section~\ref{sec:experiments}.}
    \label{table:available-methods}
\end{table*}

\subsection{Application to EELS and feasibility}

The previous subsection focused on the related works in microscopy which are summarized in Table~\ref{table:available-methods}. They are rated depending on their computational complexity and accuracy and grouped into three main families: NN, LS-regularized and DL-based methods. The works from the literature related to each method are given and separated depending on their ability to reconstruct 2D  mono-band images (e.g., HAADF) or 3D spectrum-images (e.g., EELS).


Among the available methods for EELS reconstruction, NN is fast but gives generally poor reconstruction results while DL-based methods are very efficient but remains computationally expensive, especially when considering 3D patches. 
Therefore, this literature review shows that none of the available methods can optimally reconstruct a spatially sub-sampled spectrum-image fast enough to be included into an experimental setup for online or mini-batch processing. 
Note that 3S could satisfy the accuracy and speed requirements but the results of Section~\ref{sec:experiments} will show its regularization is not suited for atomic-scale images. 
%
In this work, we will propose \corr{to apply BPFA to EELS images, i.e., recovering a dictionary composed of 3D patches, as this was originally designed for}.

An alternative for systematically reconstructing a spectrum-image consists in processing \emph{separately} and \emph{in parallel} the 2D images associated with each channel. In this case, note that NN as a 3D reconstruction performs the same as a band-by-band processing. Yet, this is sub-optimal as the reconstruction task is expected to perform better by capitalizing on the information of the whole 3D data. Similar considerations could lead to prefer reconstructing one or several single channel images of interest necessary for element mapping. However such a strategy may be also sub-optimal when no a priori knowledge is available regarding the sample to be imaged. 

To conclude, NN and DL-based methods are not suited for on-line reconstruction and only LS-regularized methods combine accuracy and reduced computational cost. As a consequence this paper proposes a method which belongs to the regularized LS family to reconstruct \corr{quickly and efficiently} an atomic-scale spectrum-image. This method is detailed in the next section and will be compared to existing approaches in Section~\ref{sec:experiments} based on synthetic as well as real data experiments.

%% file: sections/3-proposed-method.tex

\section{Proposed method}\label{sec:proposed-method}

\subsection{Structured sparsity in a well-chosen basis}

To get a fast and accurate reconstruction technique, LS-regularized methods exploiting sparsity in an appropriate basis seems to provide a relevant trade-off. \corr{In particular, to ensure an acceptable computation time, this basis or dictionary can be chosen beforehand, exploiting some expected properties of the image to be reconstructed. Learning this dictionary during a pre-processing step would be time consuming and the reconstruction results would strongly depend on the training set.} As a consequence, an analytic basis such as Fourier, DCT or wavelets will be preferred due to its explicit expression. The choice of the basis will be discussed in Section~\ref{subsec:bestbasis} based on experiments. Yet, the periodic structure of atomic-scale sample is expected to favor Fourier or DCT representations.

\begin{figure}[htbp]
    \centering
    \newlength\StrSparsityLength
    \setlength\StrSparsityLength{0.4\columnwidth}
    \begin{tabular}{ccc}
        &Image&Thresh. DCT\\
        \rotatebox{90}{HAADF}&
        \includegraphics[width=\StrSparsityLength]{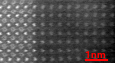}&
        \includegraphics[width=\StrSparsityLength]{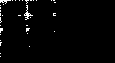}\\[20pt]
        \rotatebox{90}{Band \#1047}&
        \includegraphics[width=\StrSparsityLength]{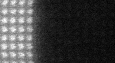}&
        \includegraphics[width=\StrSparsityLength]{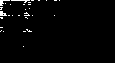}\\[20pt]
        \rotatebox{90}{Band \#1111}&
        \includegraphics[width=\StrSparsityLength]{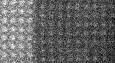}&
        \includegraphics[width=\StrSparsityLength]{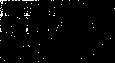}\\[20pt]
        \rotatebox{90}{Band \#1451}&
        \includegraphics[width=\StrSparsityLength]{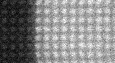}&
        \includegraphics[width=\StrSparsityLength]{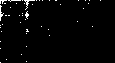}\\
    \end{tabular}

    \caption{Illustration of the data structured sparsity. The STEM acquisition is composed of a 2D HAADF image (left panel, $1$st row) and the corresponding spectrum-image whose three bands are considered (left panels, $2$nd to $4$th rows). For each image, the locations of the 2\% DCT coefficients of highest magnitude are depicted (right panels). Note that most of these coefficients are located in the same areas of the DCT space for the HAADF as well as corresponding bands, \corr{which suggests a certain structured sparsity in the the transform domain}.}
    \label{fig:joint-sparsity}
\end{figure}

More precisely, in the case of atomic-scale EELS spectrum-images, each 2D image in a given band is expected to exhibit a periodic pattern which can be described thanks to a sparse representation in an appropriate basis. Conversely, each spectrum measured in any spatial location does not likely exhibit particular periodicity. Thus this sparsity property only holds in the \emph{spatial} direction of the 3D datacube and a band-by-band basis transformation is expected to lead to higher sparsity level (i.e., the proportion of nonzero coefficients) compared to 3D basis transformation.
Besides, band-by-band representations are expected to share common characteristics since the spatial structure is likely the same across the channel and only depends on the sample. This \emph{structured sparsity} tends to promote images which non-zero band-by-band representation coefficients are located at the same place. This phenomenon is illustrated in Fig.~\ref{fig:joint-sparsity} when considering a DCT representation (the choice of the basis will be discussed in Section \ref{subsec:bestbasis}). Note that in the case of an important noise level (such as for the band \#1111), some powerful high-frequency coefficients correspond to noise (at the bottom-right corner of the panels). Yet, by jointly analyzing the DCT representations for several channels, the location of the main coefficients can be estimated after removing coefficients associated with noise.

\subsection{\CLS{0} reconstruction}\label{subsec:FS-method}
Capitalizing on the finding of the previous sections, the proposed method relies on a regularized LS problem whose formulation writes
\begin{equation} \label{eq:LS}
\hat{\X} = \arg\min_{\X}\frac{1}{2}||\Y - \X\Phi||_\mathrm{F}^2 + \lambda \mathcal{R}(\X).
\end{equation}
where $\hat{\X}$ is the recovered image, $\Phi$ is the subsampling operator and \Y\ is the acquired data. The operator $\mathcal{R}(\cdot)$ is a regularization and $\lambda$ is a scalar that adjusts the importance of the regularization with respect to the data fidelity term.
To enforce the periodic spatial patterns to be similar across the channels, \corr{we choose the regularization $\mathcal{R}(\X)$ to be}
\begin{equation}\label{eq:regularization}
\mathcal{R}(\X) = ||\X\Psi||_{2, 1}
\end{equation}
\corr{where $\Psi$ is an orthogonal band-by-band basis transform. The $\ell_{2, 1}$-norm is defined as\footnote{\corr{A slightly different definition of the $\ell_{2, 1}$-norm is proposed in~\cite{kowalski2009sparse}, where the $\ell_2$-norm is first computed on the rows of $\U$.}}}
\begin{equation}
||\U||_{2, 1} = \sum_j||\U_j||_2 = \sum_j \sqrt{\sum_i |\U_{ij}|^2}
\end{equation}
\corr{where $\U_j$ is the $j$th column of \U, corresponding to the spectrum associated with the $j$th pixel. Minimizing this norm enforces the structured sparsity as it aims at setting the columns of lowest magnitude to zero while preserving the most powerful columns. In case of a periodic basis (e.g., Fourier or DCT), the regularization \eqref{eq:regularization} promotes similar and sparse frequency representations across all the bands. For further details, see Appendix~\ref{appendix:FS}. In the sequel of the paper, the resulting method will be referred to as \emph{\CLS{0}} (\CLS{}).}


\CLS{} can be used directly on the partially-acquired spectrum-image. Yet, as a preprocessing step, \corr{we propose to perform a principal component analysis (PCA) of the acquired data $\Y$ and to apply \CLS{} on the $T$ first principal components. The formulation \eqref{eq:LS} remains the same but the measurement matrix \Y\ and reconstructed image matrix $\X$ are replaced by the PCA spectrum-image \Yt\ and its reconstructed counterpart $\Xt$. The fully reconstructed image $\X$ can then be derived from $\Xt$ by the corresponding inverse transformation. While it is not a prerequisite of the proposed method, this preprocessing step has several advantages. First, it implicitly introduces a spectral regularization of the inverse problem by imposing a low-rank structure of the solution. Similar strategies have been widely promoted for various tasks conducted on multiband images, including compressive sensing \cite{Zhang2011,Martin2014hyca}, inpainting \cite{monier2018tci,Shuang2018}, fusion \cite{Wycoff2013,Simoes2014b,Wei2015} or mixture analysis \cite{Dobigeon2009,Dobigeon2012ultra}. Second, by reducing the amount of data to process, it allows the computational time to be significantly reduced. This preprocessing may induce some reconstruction artefacts when the estimation of the covariance matrix is not accurate, e.g., in case of a low sampling ratio. Moreover, the \corr{number of PCA components to keep (which will be referred to as \emph{PCA threshold} in the following)} should be carefully chosen as a too small threshold can make some low-powerful structures disappear~\cite{mevenkamp2017mm}. A systematic strategy to adjust this threshold was proposed in \cite{monier2018tci}. The use of PCA as a pre-processing step is discussed in the supporting document~\cite{Monier2020suppNum}.}


\corr{Besides, the choice of the regularizing parameter $\lambda$ in \eqref{eq:LS} is non-trivial in general inverse problems. However, it can be fairly adjusted by exploiting some information about the image to be reconstructed such as the noise level (which can be estimated beforehand) or the sparsity level, i.e., the ratio of non-zero coefficients of the representation. This prior knowledge can be used to assess the quality of the solution $\hat{\X}_{\lambda}$ obtained by CLS for a given value $\lambda$ of the regularization parameter. For instance, for a relevant solution, the data fidelity term is expected to be of the order of magnitude of the noise level. Thus a dichotomic search can be conducted to adjust the regularization parameter automatically. In other words, \CLS{} is run for a given value of the parameter and the data-fidelity term is evaluated at convergence. Since the fidelity term increases with $\lambda$, if its value is below (resp. above) the noise level, $\lambda$ should be increased (resp. decreased) and \CLS{} should be run again. For instance, a similar strategy has been successfully followed in \cite{monier2018tci}.}
%


\subsection{\CLS{} as a pre-processing step}

As it will be shown in Section \ref{sec:experiments}, the \CLS{} algorithm is fast and it efficiently reconstructs spectrum-images based on partially acquired data. Thus we also propose here to apply \CLS{} as a pre-processing step for more advanced algorithms, in particular to initialize DL-based methods. 
To that end, the \CLS{}-reconstructed data $\hat{\X}$ is decomposed into a dictionary and a sparse representation using the conventional mini-batch DL algorithm proposed in~\cite{mairal2009online} coupled to the orthogonal matching pursuit~\cite{pati1993orthogonal}. These dictionary and code are subsequently used to initialize DL-based reconstruction methods. 
This initialization is particularly interesting for wKSVD and ITKrMM which solve a nonconvex optimization problem based on a direction alternating scheme. Indeed, for such nonconvex problems, initialization is known to be a crucial issue to ensure convergence to a relevant solution. 
Alternatively, a similar strategy cannot be easily adopted to initialize BPFA since it implements a Markov chain Monte Carlo algorithm and the dictionary and code distributions depend on hyperparameter distributions. 
The relevance of \CLS{} as an initialization step will be discussed in Section~\ref{sec:FS-init-experiments}.

%% file: sections/4-experiments.tex
\begin{table*}
    \centering
    \resizebox{\textwidth}{!}{
    \setlength\mycolwidth{2.5cm}
    \input{table/samples.tex}
    }
    \caption{Additional information about the $\R_1$, $\R_2$ and $\S$ images.}
    \label{table:samples}
\end{table*}

\section{Experiments}\label{sec:experiments}

\subsection{Data}\label{subsec:exp-data}

\subsubsection{Materials}\label{subsubsec:material-method}

Two real atomic-scale spectrum-images referred to as  $\R_1$ and $\R_2$ have been acquired on a NION UltraSTEM200 at the Laboratoire de Physique des Solides (LPS), Orsay, France. $\R_1$ is a spectrum-image acquired at $100$kV on a $\mathrm{La}_{1‐\mathrm{x}}\mathrm{Sr}_{\mathrm{x}}\mathrm{MnO}_3$/$\mathrm{Pb}(\mathrm{Zr,Ti})\mathrm{O}_3$ (LSMO/PZT) heterostructure grown onto $\mathrm{SrTiO}_3$~\cite{li2016charge}. The spectra were acquired in an energy range corresponding to $\mathrm{Ti-L}_{2,3}$, $\mathrm{O-K}$, $\mathrm{Mn -L}_{2,3}$ and $\mathrm{La-M}_{4,5}$ edges. $\R_2$ is a spectrum-image acquired at $100$kV on a $\mathrm{NdNiO}_3$ thin film grown on a $\mathrm{LaAlO}_3$ substrate~\cite{preziosi2018direct}. The spectra were acquired in an energy range corresponding to $\mathrm{O-K}$, $\mathrm{La-M}_{4,5}$, $\mathrm{Ni L}_{2,3}$ and $\mathrm{Nd-M}_{4,5}$ edges. To \corr{reduce the acquisition noise}, the images were acquired with relatively long dwell time. Additional information about the\corr{se} images (such as their sizes and resolutions, dwell times and the PCA threshold used in Section~\ref{subsec:synthetic-results}) are reported in Table~\ref{table:samples}. 
%

All experiments discussed in this section were conducted on an Intel Xeon CPU E5540 @ 2.53GHz with 8 cores -- including hyperthreading -- and 50Gb of memory. Note that the high amount of memory is only required for BPFA as a 3D algorithm while other methods can run on a machine with only 13.2Gb of memory.

\subsubsection{Synthetic and semi-real images}

Since $\R_1$ and $\R_2$ are naturally corrupted by noise, computing relevant metrics to assess the performance of the reconstruction methods may be biased by an unknown noise level. To alleviate, noise-free counterparts of $\R_1$ and $\R_2$ were first generated by conduct\corr{ing} a PCA and keeping only the first $T$ principal components. The choice of the threshold $T$ is generally a difficult task. In this work, it was set such that the discarded principal components did not carry any spatial information. More details are provided in Appendix~\ref{appendix:synth-details}. These denoised images, referred to as $\bar{\R}_1$ and $\bar{\R}_2$, are assumed to be the ground-truth images \X\ to be reconstructed.

In addition to these two semi-real images, a synthetic spectrum-image was generated based on the $\R_2$ data. To that end, an independent component analysis was conducted on $\R_2$ to extract four characteristic spectra that were filtered afterwards while the associated spectral maps were synthetically produced. These data were subsequently mixed to get the spectrum-image referred to as $\S$ (see Appendix~\ref{appendix:synth-details} for generation details).

To mimic realistic experimental setups, the three noise-free images $\bar{\R}_1$,  $\bar{\R}_2$ and $\S$ were subsequently corrupted by an additive white Gaussian noise with signal-to-noise ratio adjusted in agreement with the acquisition process. Finally these pseudo-real images, referred to as $\R_1^*$ and $\R_2^*$, and the noisy synthetic image $\S^*$, were \corr{uniformly} randomly spatially subsampled with a 20\% ratio to provide the measurement matrix $\Y$. {Note that results obtained for other sampling ratios are reported in the supporting document \cite{Monier2020suppNum}.}

\subsection{Methods}

As discussed in Section \ref{subsec:FS-method}, for all algorithms, a PCA is first conducted on the observed image, keeping the $T$ most relevant principal components. The reconstruction algorithms are then run in this lower-dimensional subspace. The inverse transformation will be applied afterwards to get the reconstructed image $\hat{\X}$. Compared methods are NN, 3S, \CLS{}, ITKrMM, wKSVD and BPFA. In particular, NN is the only one to be applied band-by-band. 
%
\me{For all methods, the algorithmic parameters have been adjusted to reach the best performances for each method. In particular, }
DL-based methods consider 3D \corr{patches} of size $M\times M \times T$ with $M=25$ for ITKrMM and wKSVD and $M=41$ for BPFA.

\corr{
 The ITKrMM and wKSVD implementations used in these experiments are the Matlab codes provided by Prof. K. Schnass\footnote{\corr{\url{https://www.uibk.ac.at/mathematik/personal/schnass/code/itkrmm.zip}}}. The implementation of BPFA is the Matlab code provided by Dr. Z. Xing\footnote{\corr{\url{https://drive.google.com/open?id=0B9548VKFKtmiY2ZNRFVUTjhyUFE}}}. The other methods have been implemented by the authors of this paper and are available in a Python library called \texttt{inpystem}\footnote{\corr{\url{https://github.com/etienne-monier/inpystem}}}. The codes to reproduce the experiments described in this paper are also available online\footnote{\corr{\url{https://github.com/etienne-monier/2020-Ultramicro-fast}}}.
}


\subsection{Metrics}

To evaluate the reconstruction quality, \corr{several} quantitative measures will be used to compare the ground truth \X\ and reconstructed  $\hat{\X}$ images. \corr{First, the normalized mean-squared error (NMSE) is chosen as an error measure and is  computed according to
\begin{equation}\label{eq:NMSE}
    \mathrm{NMSE}(\hat{\X},\X) = \frac{||\hat{\X}-\X||_{\mathrm{F}}^2}{||\X||_{\mathrm{F}}^2}.
\end{equation}
The smaller NMSE, the better. Then, this error measure is turned out as a performance measure by considering its negative-logarithm, defining the signal-to-noise ratio (SNR)
\begin{equation}\label{eq:SNR}
    \mathrm{SNR}(\hat{\X},\X) = -10\log_{10}\left({\mathrm{NMSE}(\hat{\X},\X)}\right). 
\end{equation}}
The higher SNR, the better. 
Additionally, we also consider the average spectral angle distance (aSAD) defined as~\cite{keshava2004distance,sohn2002supervised}
\begin{equation}
    \mathrm{aSAD}(\hat{\X},\X) = \frac{1}{\P}\sum_{j=1}^{\P}\mathrm{acos}\left(
        \frac
        {\langle\hat{\X}_j, \X_j\rangle}
        {||\hat{\X}_j||_2\times||\X_j||_2}
    \right),
\end{equation}
where \P\ is the number of pixels. The aSAD, which is a measure of spectral distorsion, is independent of scaling and should be close to zero.
%
\corr{Finally, the structural similarity index (SSIM) averaged over all the bands is considered as a criterion to assess spatial reconstruction~\cite{wang2009mean,wang2004image}. A value close to 1 (resp. to 0) indicates that the spatial structures are similar (resp. different). The closer to 1, the better.}

%% file: table/samples.tex
\bgroup
    \renewcommand{\arraystretch}{1.2}
    \begin{tabular}{%
            c%
            >{\centering\arraybackslash}m{\mycolwidth}%
            m{1.3\mycolwidth}%
            >{\centering\arraybackslash}m{\mycolwidth}%
            c%
            c%
            >{\centering\arraybackslash}m{2cm}}%
        \toprule
        \multirow{2}*{Image}& 
        \multirow{2}*{Sample}& 
        \multirow{2}{1.3\mycolwidth}{%
            \me{Size ($x$, $y$, $\lambda$)\newline Res. ($\Delta x = \Delta y$, $\Delta \lambda$)}}&
        \multirow{2}{0.7\mycolwidth}{%
            \centering\arraybackslash Dwell time (ms)}&
        \multicolumn{2}{c}{Relevant edges}&
        \multirow{2}{2cm}{\centering\arraybackslash PCA threshold $T$}\\
        &&&&Element&Energy loss (eV)&\\
        \midrule
        \multirow{4}*{$\mathsf{R}_1$}& 
        \multirow{4}{\mycolwidth}{\centering\arraybackslash 
                      PbZrTiO\textsubscript{3} / 
                      LaSrMnO\textsubscript{3} / 
                      SrTiO\textsubscript{3}
                      }& 
        \multirow{4}{1.3\mycolwidth}{(232, 101, 1530)\newline\me{(0.055nm, 0.27eV)}}& 
        \multirow{4}*{20}& 
        Ti& 
        456& 
        \multirow{4}*{9}\\ 
        &&&&O& 532&\\
        &&&&Mn& 640&\\
        &&&&La& 832&\\
        \midrule
        \multirow{4}*{$\mathsf{R}_2$}& 
        \multirow{4}{\mycolwidth}{\centering\arraybackslash 
                      NdNiO\textsubscript{3}/
                      LaAlO\textsubscript{3}
                      }& 
        \multirow{4}{1.3\mycolwidth}{(63, 115, \corr{1505})\me{\newline(0.045nm, 0.32eV)}}& 
        \multirow{4}*{20}& 
        O& 
        532& 
        \multirow{4}*{7}\\ 
        &&&&La& 832&\\
        &&&&Ni& 855&\\
        &&&&Nd& 978&\\
        \midrule
        $\mathsf{S}$& 
        cf. $\mathsf{R}_2$& 
        (70, 120, 1435)\me{\newline(0.045nm, 0.32eV)}& 
        -& 
        \multicolumn{2}{c}{cf. $\mathsf{R}_2$}&
        4\\ 
        \bottomrule
    \end{tabular}
\egroup

%% file: sections/5-results.tex

\section{Results}\label{sec:results}

\subsection{Appropriate basis for sparsity}\label{subsec:bestbasis}

As explained in the previous section, the data sparsity level highly depends on the image and on the basis chosen for its representation. In the case of atomic-scale images, previous works considered wavelets~\cite{li2018compressive}, Fourier~\cite{stevens2018apl} or DCT~\cite{Beche2016development,anderson2013sparse} bases. However, no systematic comparison of these transforms has been conducted in these works. This section proposes to fill this gap.
Intuitively, Fourier transform or DCT are expected to provide best results as they are known to lead to highly sparse representations when applied to periodic structures. To confirm this intuition, we propose to monitor the magnitude of the data representation coefficients which are expected to decrease as fast as possible when sorted in decreasing order. A significant decreasing would mean that less coefficients are required to accurately represent the data. Thus, the reconstruction error is evaluated for each image when represented with the following transforms:
 Fourier basis, DCT basis, Daubechies and symlet wavelets (with 3, 10 and 20 vanishing moments).
The reconstruction error is evaluated as a function of ratio $r \in (0,1)$ of nonzero representation coefficients, i.e., keeping only the $r$\% representation coefficients of highest magnitude. The inverse basis transformation provides an estimate \corr{$\hat{\X}$} of \corr{\X} which is degraded as $r$ decreases. The reconstruction error correspond to the normalized mean square error between \X\ and $\hat{\X}$.
To summarize, the most suitable transform for a given image should be the one whose reconstruction error drops fastest with $r$ as this shows that less coefficients are necessary to describe the image.

\begin{figure}[htbp]
    \centering
    \input{img/approp_basis}
    \caption{Reconstruction error in term of NMSE for several bases when representing $\R_1$ (top), $\R_2$ (middle) and $\S$ (bottom). The faster the curve decreases, the better as it means the image needs less representation coefficients to be accurately represented. The DCT basis gives the best results for all images.}
    \label{fig:best-basis}
\end{figure}
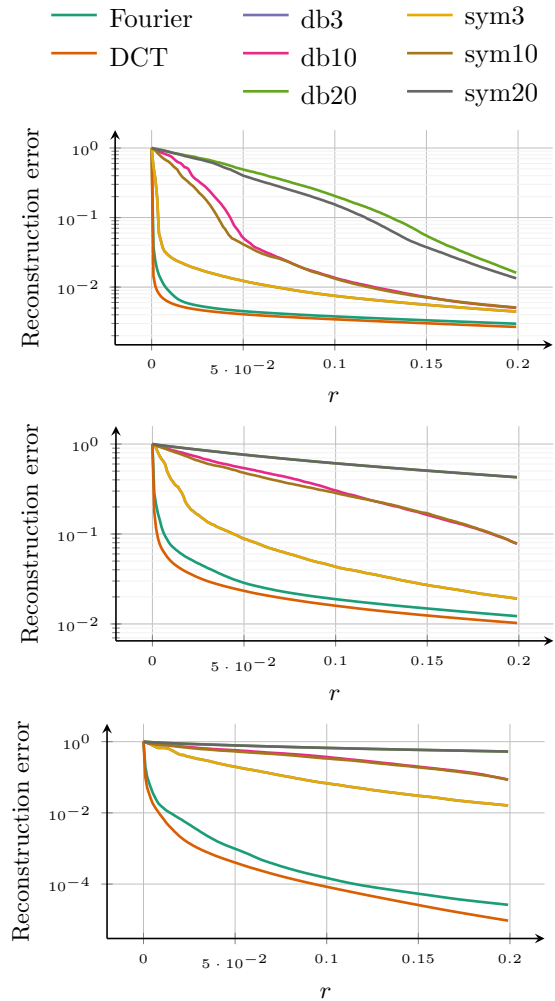

This procedure has been applied \corr{to $\R_1$, $\R_2$ and $\S$ after a PCA}. The results are depicted in Fig.~\ref{fig:best-basis}. They show that DCT is always better than Fourier as the measurements are real-valued data, which introduces a redundancy of factor 2 for each axis in the 2D Fourier space whereas DCT does not suffer from this redundancy. Moreover, our intuition is confirmed since DCT and Fourier perform better for the three images compared to all the wavelet transforms. Finally, in the case of atomic-scale images considered in this work, these experiments show that DCT offers the sparsest transform and it will be chosen as the orthogonal band-by-band transform $\Psi$ in \eqref{eq:regularization} of the proposed \CLS{} method.

\subsection{Results on synthetic and semi-real images}\label{subsec:synthetic-results}

\begin{table}[htbp]
    \centering
    \input{table/synth_results}
    \caption{Reconstruction performance in terms of SNR, aSAD and SSIM for semi-real  $\R_1^*$ and  $\R_2^*$ and synthetic  $\S^*$ images. The execution time is also given to be considered jointly with accuracy. \corr{There is a clear performance gap in terms of quality and execution time between NN and the DL-based methods . LS-regularized methods fill this gap, especially CLS which performs well in comparison to 3S.}}
    \label{table:results-time}
\end{table}

\corr{In this section, the reconstruction methods were applied to the two semi-real images $\R_1^*$ and $\R_2^*$ and to the synthetic image $\S^*$.} 
%

The metric values as well as the execution times are reported in Table~\ref{table:results-time}. These results show that there is a clear performance gap, both in terms of quality and time, between NN and DL-based techniques. This is particularly true for BPFA which exhibits a prohibitive computation time, yet often leading to the best image reconstruction. 
The LS-regularized methods 3S and \CLS{} seem to fill this gap, both in terms of accuracy and speed. \CLS{} performs particularly well as it gives a SNR close to the best methods with a very small computational time. 3S gives \corr{lower performance} results (even if better that NN) since its regularization is not appropriate to accurately describe the periodic structure of atomic-scale images. 
%
Note also that BPFA gives globally the best SNR and aSAD, except for ${\R}_1^*$ for which \CLS{}\ gives a better aSAD, while NN gives the worst aSAD. 

\begin{figure}[htbp]
    \centering
    \subfigure[Band \#2 of $\bar{\R}_2$]{\includegraphics[height=0.2\textwidth]{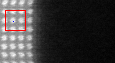}\label{subfig:location_samp}}\\
    \subfigure[Band \#2 of $\bar{\R}_2$ (zoom)]{\includegraphics[height=0.15\textwidth]{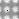}\label{subfig:location_samp-zoom}}\hskip 20pt
    \subfigure[Mask (zoom)]{\includegraphics[height=0.15\textwidth]{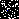}\label{subfig:location_samp-zoom-mask}}\\
    \caption{\corr{Location of the sampled spectrum represented in Figure~\ref{fig:synth-image-spectra}.} 
    Image~\ref{subfig:location_samp} shows \corr{the 2nd principal component} of the semi-real image $\bar{\R}_2$ and locates the sampled pixel (blue dot)  whose spectrum is considered in Fig. \ref{fig:synth-image-spectra}. Fig.~\ref{subfig:location_samp-zoom} and \ref{subfig:location_samp-zoom-mask} shows zooms on the image region-of-interest and on the sampling mask. The sampling mask white (resp. black) pixels stand for sampled (resp. non-sampled) pixels.}
    \label{fig:synth-image-spectra-locations}
\end{figure}

\begin{figure}[htbp]
    \centering
    \extcode{%
        \input{img/spectra_template}
    }{%
        \subfigure[Non-sampled pixel]{\includegraphics[scale=1]{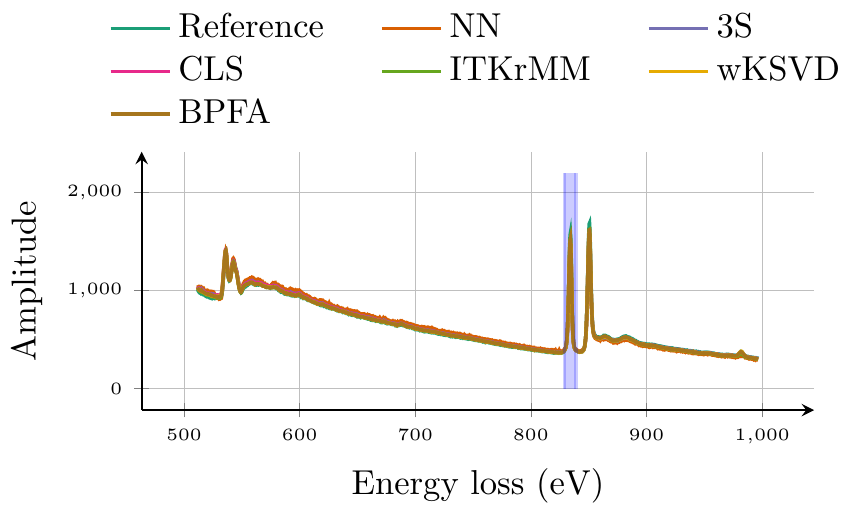}\label{subfig:synth_spectra_non_samp}}\\
        \subfigure[Non-sampled pixel (zoom)]{\includegraphics[scale=1]{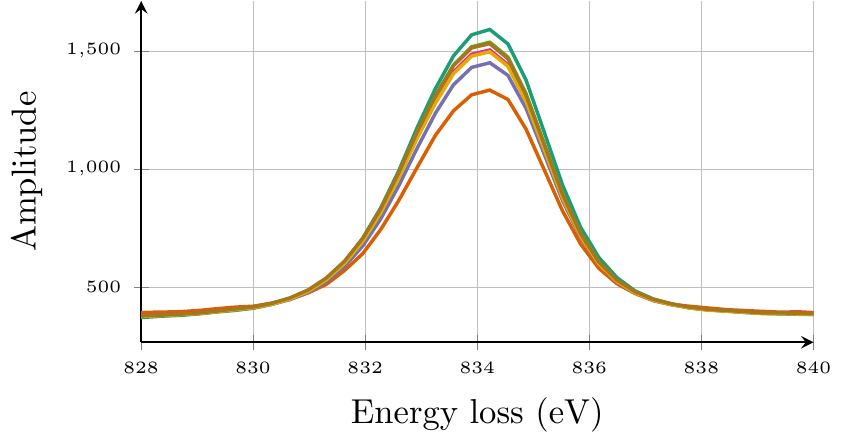}}\\
    }
    %
    \caption{Spectra reconstruction results for $\R_2^*$ for a non-sampled pixels located in Fig.~\ref{fig:synth-image-spectra-locations}. Reference spectra corresponds to the noise-free image $\bar{\R}_2$. The zoom represents the region-of-interest highlighted as shaded a blue area. \corr{NN and 3S spectra are significantly shifted compared to the reference whereas the spectra recovered by CLS and DL-based methods are close to the reference spectrum.}}
    \label{fig:synth-image-spectra}
\end{figure}

\corr{The reconstruction of a non-sampled pixel located in Fig. \ref{fig:synth-image-spectra-locations} is also depicted in Fig.~\ref{fig:synth-image-spectra}. In this figure, reference data refers to the noise-free image $\bar{\R}_2$. Equivalent plots for a sampled pixel are omitted here since they do not bring any meaningful insight: the reconstructed spectra are close to be distinguished. These plots show that the NN-reconstructed spectrum is significantly shifted with respect to the reference while BPFA and \CLS{} are close to the reference spectrum. Error maps are reported in the supporting document \cite{Monier2020suppNum}.}

As a consequence, \CLS{} appears as a relevant trade-off between accuracy and complexity since it gives good reconstruction with small computational time. This method could be interesting as an experimental tool whereas BPFA could be used as a post-processing refinement method. Combining both methods will be discussed in Section \ref{sec:FS-init-experiments}.


\subsection{Results on a real image}

Some illustrative results are also provided for $\R_2$. More precisely, the real spectrum-image $\R_2$ is spatially subsampled with a ratio of $20$\% and then reconstructed as in the previous subsection. 

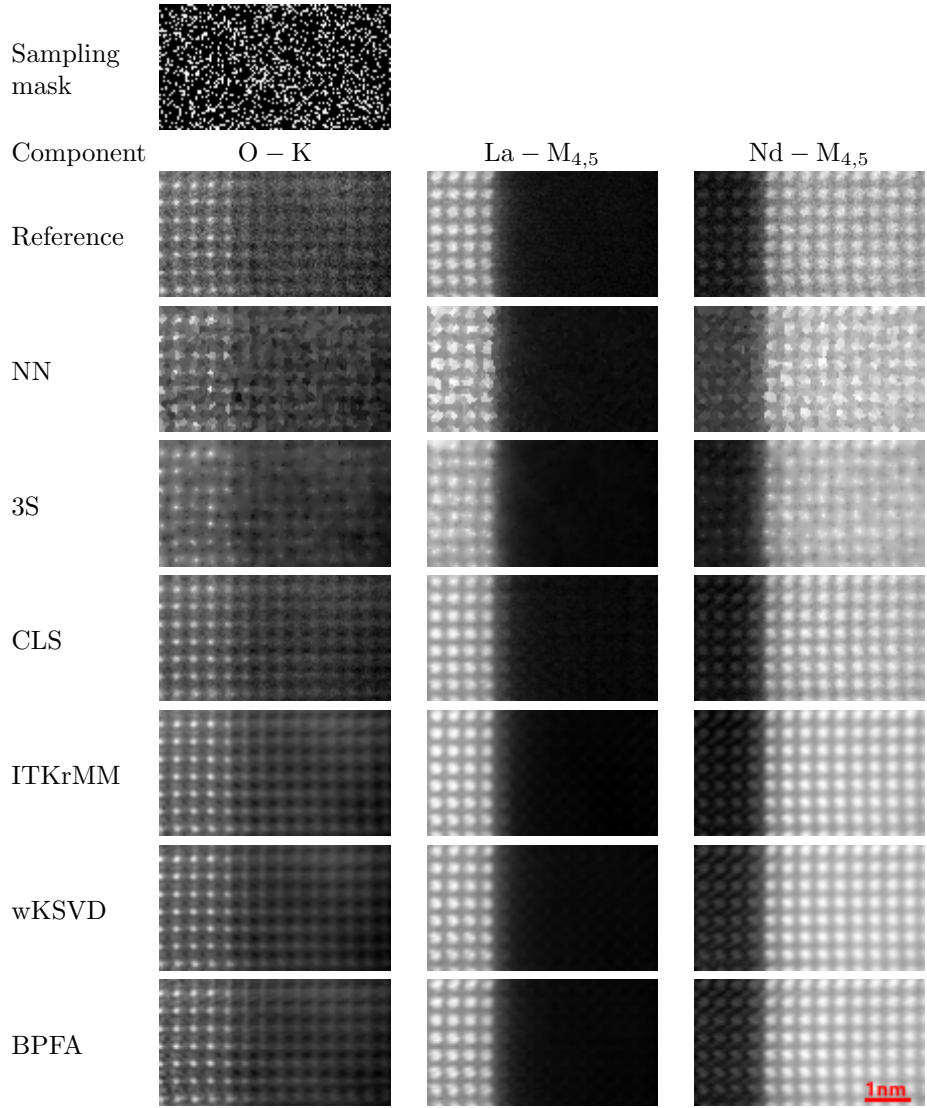
\begin{figure*}[htbp]
    \centering
    \setlength\mycolwidth{0.25\textwidth}
    \input{table/real_im_results}
    \caption{Reconstruction results for $\R_2$. The images show the sum of 5 bands around 3 particular edges ($\mathrm{O-K}$, $\mathrm{La-M}_{4, 5}$ and $\mathrm{Nd-M}_{4, 5}$). The reference corresponds to the real, possibly noisy, image ${\R}_2$. The sampling mask is also provided in the first row where white (resp. black) pixels stand for sampled (resp. non-sampled) pixels. \corr{These results confirm the performance gap between NN whose images are not smooth enough and DL-based methods which are close to the reference with an additional denoising effect. CLS performs clearly better than NN and 3S and its results are close to DL-based methods.}}
    \label{fig:results-HR2_real-bands}
\end{figure*}

\begin{figure}[htbp]
    \centering
    \extcode{%
        \input{img/spectra_template_real}
    }{%
        \subfigure[Non-sampled pixel]{\includegraphics[scale=1]{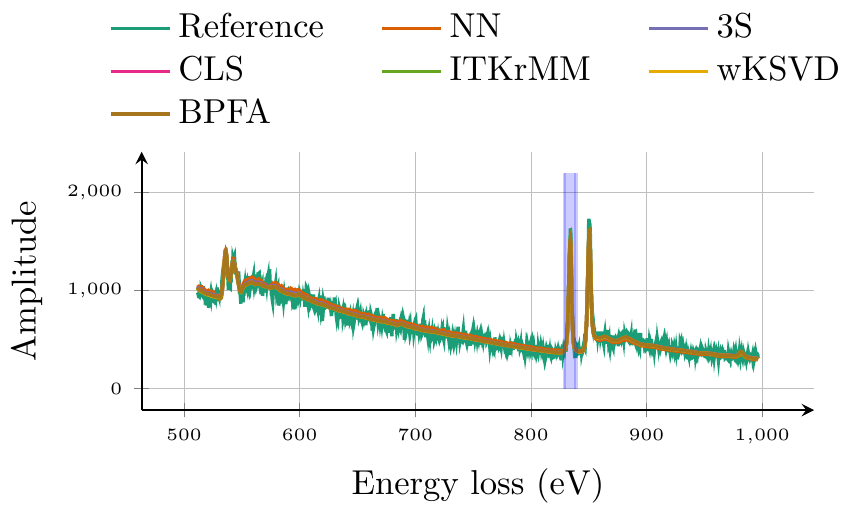}}\\
        \subfigure[Non-sampled pixel (zoom)]{\includegraphics[scale=1]{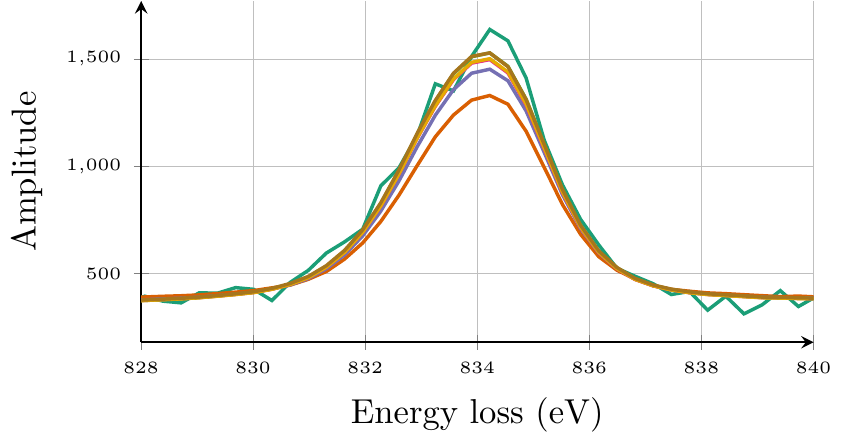}}\\
    }
    %
    \caption{Spectra reconstruction results for $\R_2$ for a non-sampled pixel with the same locations as in Fig.~\ref{fig:synth-image-spectra-locations}. Reference spectra corresponds to the real image ${\R}_2$. The zoom represent the region-of-interest highlighted as a shaded blue area. \corr{As for the synthetic results, the NN spectrum is significantly shifted compared to the reference spectrum while the results of CLS and DL-based methods (especially BPFA) are close to the reference. Again, CLS appears to be a relevant trade-off.}}
    \label{fig:real-image-spectra}
\end{figure}

\corr{Visual representations of the reconstructed spectrum-images around some interesting edges are provided in Fig.~\ref{fig:results-HR2_real-bands} and the reconstruction of a non-sampled spectrum is shown in Fig.~\ref{fig:real-image-spectra}. In these figures, note that reference data refers to the real, possibly noisy, image $\R_2$.} 

Similarly to the previous \corr{findings}, DL-based methods visually give excellent reconstructed maps when compared to other methods. Results regarding reconstructed spectra show that the NN-reconstructed spectrum for the non-sampled pixel exhibits a significant shift while other algorithms provide smaller biased results.

\subsection{\CLS{} as an initialization}\label{sec:FS-init-experiments}

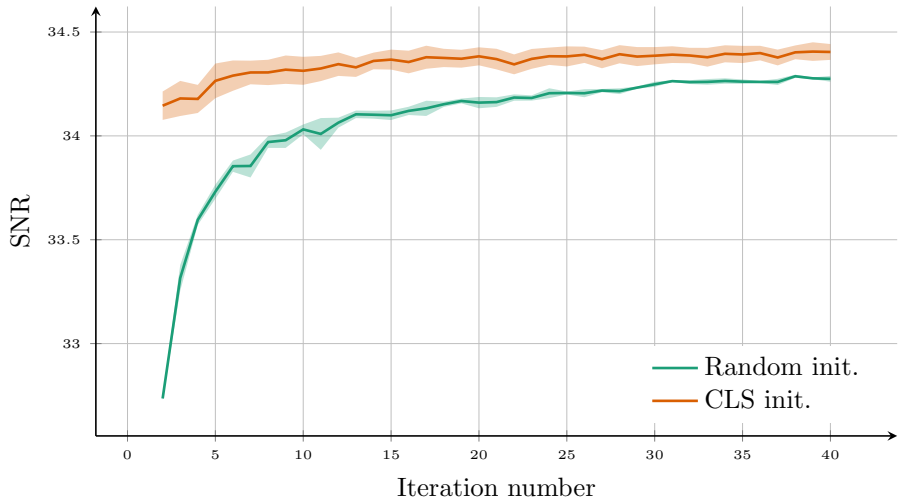
\begin{figure}[htbp]
    \centering
    \input{img/CLS_init}
    \caption{SNR as a function of wKSVD iterations for random and \CLS{} initializations. Colored filling corresponds to standard deviation interval computed from $10$ runs. \corr{The results show that CLS-based initialization needs less iterations to get a target SNR compared to random initialization. This promotes CLS as an initialization for DL-based reconstruction methods.}}
    \label{fig:FS_init}
\end{figure}

Previous subsections compared the reconstruction method performances  on synthetic, semi-real and real images. They illustrated the interest of \CLS{} as a fast and efficient method for reconstructing subsampled atomic-scale spectrum-images. Besides, DL-based reconstruction methods are shown to provide \corr{higher performances} at the price of a \corr{higher} computational cost. However, coupled with \CLS{}, these methods may be used as post-experiment refinement. In this section, we will confirm this interest based on an additional experiment conducted on the semi-real image $\R_2^*$. Similarly to the previous paragraph, this image has been subsampled and then reconstructed with wKSVD while considering  two distinct initializations. The first one is purely random, as initially implemented by the algorithm. The second kind of initialization consists in initializing wKSVD with a dictionary and a corresponding sparse code derived from the \CLS{}-based reconstructed image. Then, the reconstruction quality is monitored as a function of the iterations of the wKSVD algorithm by plotting the reconstruction SNR. Results are depicted in Fig.~\ref{fig:FS_init} for both initializations.

The results show that \CLS{}-based initialization leads to less iterations to get a target SNR compared to random initialization. It confirms the interest of \CLS{} to be used as an initialization to accelerate more elaborated yet more computationally intensive techniques. Note that this strategy can be adopted for gradient-descent-based algorithms such as wKSVD or ITKrMM, but is not suited for Markov chain Monte-Carlo algorithms such as BPFA.

%% file: img/approp_basis.tex
\extcode{%
    \begin{tikzpicture}[ampersand replacement=\&]
        \begin{axis}[
            ymode=log,
            xlabel={$r$},
            label style={font=\small},
            width=0.4\textwidth,
            height=0.5\columnwidth,
            ylabel={Reconstruction error},
            name=aa,
            ]
            \addplot+ [line width=1pt] table [x=rTab, y=fourier]{img/2.basis/HR1.dat};
            \label{plot:Fourier}
            \addplot+ [line width=1pt] table [x=rTab, y=dct]{img/2.basis/HR1.dat};
            \label{plot:DCT}
            \addplot+ [line width=1pt] table [x=rTab, y=db3]{img/2.basis/HR1.dat};
            \label{plot:db3}
            \addplot+ [line width=1pt] table [x=rTab, y=db10]{img/2.basis/HR1.dat};
            \label{plot:db10}
            \addplot+ [line width=1pt] table [x=rTab, y=db20]{img/2.basis/HR1.dat};
            \label{plot:db20}
            \addplot+ [line width=1pt] table [x=rTab, y=sym3]{img/2.basis/HR1.dat};
            \label{plot:sym3}
            \addplot+ [line width=1pt] table [x=rTab, y=sym10]{img/2.basis/HR1.dat};
            \label{plot:sym10}
            \addplot+ [line width=1pt] table [x=rTab, y=sym20]{img/2.basis/HR1.dat};
            \label{plot:sym20}
        \end{axis}
        \matrix [
            matrix of nodes,
            nodes={anchor=west},
            anchor=south,
            at={([shift={(-15pt,5pt)}]aa.north)},
            draw=none,
            inner sep=2pt,
            row sep=2pt
            ] {
            \ref{plot:Fourier}  \& Fourier   \&[0.5cm] \ref{plot:db3}  \& db3  \&[0.5cm] \ref{plot:sym3}  \& sym3 \\
            \ref{plot:DCT}      \& DCT       \&        \ref{plot:db10} \& db10 \&        \ref{plot:sym10} \& sym10 \\
                                \&           \&        \ref{plot:db20} \& db20 \&        \ref{plot:sym20} \& sym20 \\
            };
    \end{tikzpicture}\\
    \begin{tikzpicture}[]
        \begin{axis}[
            ymode=log,
            xlabel={$r$},
            label style={font=\small},
            width=0.4\textwidth,
            height=0.5\columnwidth,
            ylabel={Reconstruction error},
            line width=1pt,
            ]
            \addplot+ [] table [x=rTab, y=fourier]{img/2.basis/HR2.dat};
            \addplot+ [] table [x=rTab, y=dct]{img/2.basis/HR2.dat};
            \addplot+ [] table [x=rTab, y=db3]{img/2.basis/HR2.dat};
            \addplot+ [] table [x=rTab, y=db10]{img/2.basis/HR2.dat};
            \addplot+ [] table [x=rTab, y=db20]{img/2.basis/HR2.dat};
            \addplot+ [] table [x=rTab, y=sym3]{img/2.basis/HR2.dat};
            \addplot+ [] table [x=rTab, y=sym10]{img/2.basis/HR2.dat};
            \addplot+ [] table [x=rTab, y=sym20]{img/2.basis/HR2.dat};
        \end{axis}
    \end{tikzpicture}\\
    \begin{tikzpicture}[]
        \begin{axis}[
            ymode=log,
            xlabel={$r$},
            label style={font=\small},
            width=0.4\textwidth,
            height=0.5\columnwidth,
            ylabel={Reconstruction error},
            line width=1pt,
            ]
            \addplot+ [] table [x=rTab, y=fourier]{img/2.basis/Synth.dat};
            \addplot+ [] table [x=rTab, y=dct]{img/2.basis/Synth.dat};
            \addplot+ [] table [x=rTab, y=db3]{img/2.basis/Synth.dat};
            \addplot+ [] table [x=rTab, y=db10]{img/2.basis/Synth.dat};
            \addplot+ [] table [x=rTab, y=db20]{img/2.basis/Synth.dat};
            \addplot+ [] table [x=rTab, y=sym3]{img/2.basis/Synth.dat};
            \addplot+ [] table [x=rTab, y=sym10]{img/2.basis/Synth.dat};
            \addplot+ [] table [x=rTab, y=sym20]{img/2.basis/Synth.dat};
        \end{axis}
    \end{tikzpicture}
}{%
    \includegraphics{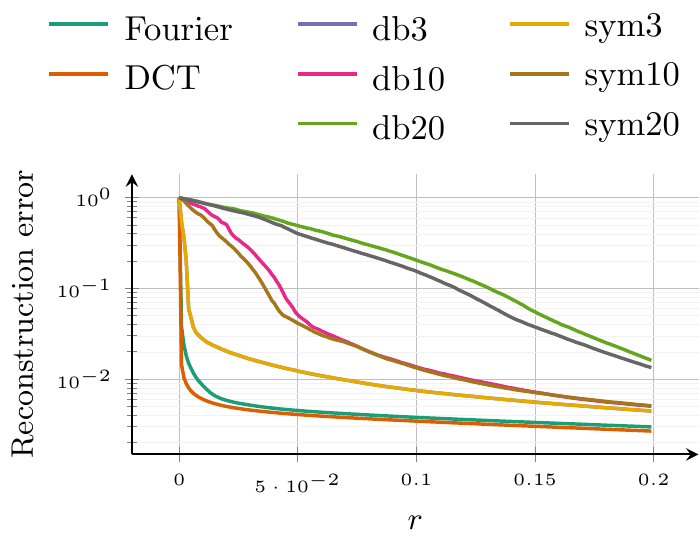}\\
    \includegraphics{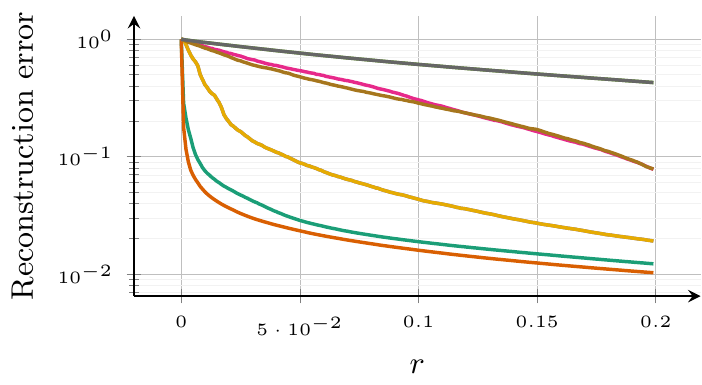}\\
    \includegraphics{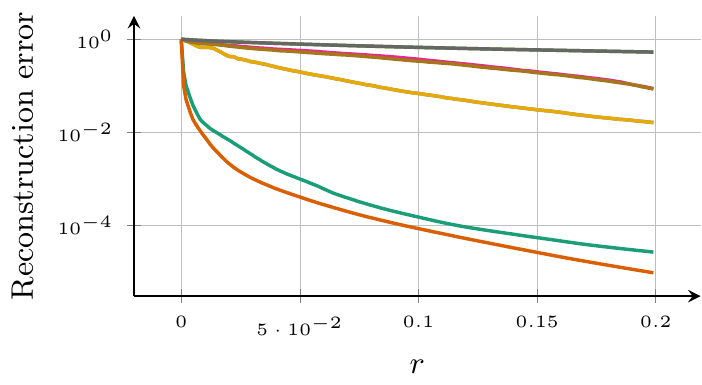}
}

%% file: table/synth_results.tex
\bgroup
    \renewcommand{\arraystretch}{1.2}
    
    \subfigure[ $\mathsf{R}_1^*$]{
        \begin{tabular}{%
                c%
                c%
                c%
                c%
                c}%
            \toprule
            Method& 
            SNR&
            aSAD (100$\times$)&
            \corr{SSIM}&
            Time(s)\\
            \midrule
	NN & 30.81 & 1.384 & 0.680 & \textbf{0.423}\\
	3S & 32.30 & 1.080 & 0.643 & 15.2\\
	CLS & 36.18 & \textbf{1.061} & 0.912 & 18.4\\
	ITKrMM & 36.52 & 1.097 & 0.923 & 6.85e+04\\
	wKSVD & 36.93 & 1.091 & 0.931 & 7.97e+04\\
	BPFA & \textbf{37.02} & 1.089 & \textbf{0.933} & 1.36e+05\\
            \bottomrule
        \end{tabular}
    }
    \subfigure[ $\mathsf{R}_2^*$]{
        \begin{tabular}{%
                c%
                c%
                c%
                c%
                c}%
            \toprule
            Method& 
            SNR&
            aSAD (100$\times$)&
            \corr{SSIM}&
            Time(s)\\
            \midrule
	NN & 28.71 & 1.815 & 0.635 & \textbf{7.94e-02}\\
	3S & 30.17 & 1.496 & 0.621 & 3.38\\
	CLS & 33.15 & 1.233 & 0.790 & 3.09\\
	ITKrMM & 33.67 & 1.253 & 0.819 & 9.55e+03\\
	wKSVD & 34.52 & 1.163 & 0.841 & 2.59e+04\\
	BPFA & \textbf{35.01} & \textbf{1.106} & \textbf{0.852} & 6.18e+04\\
            \bottomrule
        \end{tabular}
    }
    \subfigure[ $\mathsf{S}^*$]{
        \begin{tabular}{%
                c%
                c%
                c%
                c%
                c}%
            \toprule
            Method& 
            SNR&
            aSAD (100$\times$)&
            \corr{SSIM}&
            Time(s)\\
            \midrule
	NN & 21.32 & 1.462 & 0.735 & \textbf{6.82e-02}\\
	3S & 22.12 & 1.174 & 0.710 & 3.33\\
	CLS & 42.14 & 0.224 & 0.997 & 1.48\\
	ITKrMM & 44.16 & 0.338 & 0.998 & 9.61e+03\\
	wKSVD & 45.59 & 0.277 & 0.999 & 1.57e+04\\
	BPFA & \textbf{52.70} & \textbf{0.150} & \textbf{1.000} & 4.06e+04\\
            \bottomrule
        \end{tabular}
    }
\egroup

%% file: img/spectra_template.tex
\setlength{\myfigurewidth}{0.95\columnwidth}
%
%
\def\fileplot{img/3.synth_im/synth_HR2_samp.dat}
\subfigure[Sampled pixel (blue pixel)]{
    \begin{tikzpicture}[]
        
        \begin{axis}[
            xlabel={Energy loss (eV)},
            ylabel={Amplitude},
            legend entries={Reference, NN, 3S, \CLS{}, ITKrMM, wKSVD, BPFA},%
            width=\myfigurewidth,
            height=0.5\myfigurewidth,
            ymin=0, ymax=2200,
            %
            legend style={
                draw=none, 
                at={(0.5,1.03)}, 
                anchor=south, 
                legend columns=3, 
                /tikz/every even column/.append style={column sep=0.5cm}
            },
            line width=1pt,
            ]
            \addplot+ [] table [x=eV, y=GT]{\fileplot};
            \addplot+ [] table [x=eV, y=NN]{\fileplot};
            \addplot+ [] table [x=eV, y=SSS]{\fileplot};
            \addplot+ [] table [x=eV, y=FS3D]{\fileplot};
            \addplot+ [] table [x=eV, y=ITKrMM]{\fileplot};
            \addplot+ [] table [x=eV, y=wKSVD]{\fileplot};
            \addplot+ [] table [x=eV, y=BPFA]{\fileplot}; 
            
            \addplot+ [name path=llim, blue, opacity=0.2] coordinates {(829, 0) (829, 2200)};
            \addplot+ [name path=rlim, blue, opacity=0.2] coordinates {(839, 0) (839, 2200)};
            \addplot+ [blue, opacity=0.2] fill between[of=llim and rlim];
        \end{axis}
        
    \end{tikzpicture}
    \label{subfig:synth_spectra_samp}
}\\
%
%
\subfigure[Sampled pixel (blue pixel) - zoom]{
    \begin{tikzpicture}[]
        
        \begin{axis}[
            xlabel={Energy loss (eV)},
            ylabel={Amplitude},
            width=\myfigurewidth,
            height=0.6\myfigurewidth,
            xmin=829,
            xmax=839,
            line width=0.5pt,
            ]
            \addplot+ [] table [x=eV, y=GT]{\fileplot};
            \addplot+ [] table [x=eV, y=NN]{\fileplot};
            \addplot+ [] table [x=eV, y=SSS]{\fileplot};
            \addplot+ [] table [x=eV, y=FS3D]{\fileplot};
            \addplot+ [] table [x=eV, y=ITKrMM]{\fileplot};
            \addplot+ [] table [x=eV, y=wKSVD]{\fileplot};
            \addplot+ [] table [x=eV, y=BPFA]{\fileplot}; 
        \end{axis}
    \end{tikzpicture}
    \label{subfig:synth_spectra_samp_zoom}
}\\
%
%
\def\fileplot{img/3.synth_im/synth_HR2_non_samp.dat}
\subfigure[Non-sampled pixel (red pixel)]{
    \begin{tikzpicture}[]
        \begin{axis}[
            xlabel={Energy loss (eV)},
            ylabel={Amplitude},
            width=\myfigurewidth,
            line width=1pt,
            legend entries={Reference, NN, 3S, \CLS{}, ITKrMM, wKSVD, BPFA},
            legend style={
                draw=none, 
                at={(0.5,1.03)}, 
                anchor=south, 
                legend columns=3, 
                /tikz/every even column/.append style={column sep=0.5cm}
            },
            height=0.5\myfigurewidth,
            ymin=0, ymax=2200,
            ]
            \addplot+ [] table [x=eV, y=GT]{\fileplot};
            \addplot+ [] table [x=eV, y=NN]{\fileplot};
            \addplot+ [] table [x=eV, y=SSS]{\fileplot};
            \addplot+ [] table [x=eV, y=FS3D]{\fileplot};
            \addplot+ [] table [x=eV, y=ITKrMM]{\fileplot};
            \addplot+ [] table [x=eV, y=wKSVD]{\fileplot};
            \addplot+ [] table [x=eV, y=BPFA]{\fileplot}; 
            
            \addplot+ [name path=llim, blue, opacity=0.2] coordinates {(829, 0) (829, 2200)};
            \addplot+ [name path=rlim, blue, opacity=0.2] coordinates {(839, 0) (839, 2200)};
            \addplot+ [blue, opacity=0.2] fill between[of=llim and rlim];
        \end{axis}
    \end{tikzpicture}
    \label{subfig:synth_spectra_non_samp}
}\\ 
%
%
\subfigure[Non-sampled pixel (red pixel) - zoom]{
    \begin{tikzpicture}[]
        
        \begin{axis}[
            xlabel={Energy loss (eV)},
            ylabel={Amplitude},
            width=\myfigurewidth,
            height=0.6\myfigurewidth,
            xmin=829,
            xmax=839,
            line width=0.5pt,
            ]
            \addplot+ [] table [x=eV, y=GT]{\fileplot};
            \addplot+ [] table [x=eV, y=NN]{\fileplot};
            \addplot+ [] table [x=eV, y=SSS]{\fileplot};
            \addplot+ [] table [x=eV, y=FS3D]{\fileplot};
            \addplot+ [] table [x=eV, y=ITKrMM]{\fileplot};
            \addplot+ [] table [x=eV, y=wKSVD]{\fileplot};
            \addplot+ [] table [x=eV, y=BPFA]{\fileplot}; 
        \end{axis}
    \end{tikzpicture}
    \label{subfig:synth_spectra_non_samp_zoom}
}\\

%% file: table/real_im_results.tex
\begin{tabular}{m{1.5cm}*{3}{>{\centering\arraybackslash}m{1.02\mycolwidth}}}
    Sampling mask&
    \includegraphics[width=\mycolwidth]{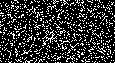}&
    &
    \\
     Component&
    {$\mathrm{O-K}$}&
    {$\mathrm{La-M}_{4, 5}$}&
    {$\mathrm{Nd-M}_{4, 5}$}\\
    Reference&
    \includegraphics[width=\mycolwidth]{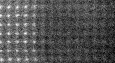}&
    \includegraphics[width=\mycolwidth]{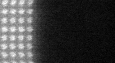}&
    \includegraphics[width=\mycolwidth]{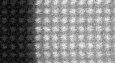}\\
    NN&
    \includegraphics[width=\mycolwidth]{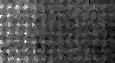}&
    \includegraphics[width=\mycolwidth]{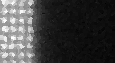}&
    \includegraphics[width=\mycolwidth]{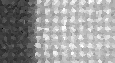}\\
    3S&
    \includegraphics[width=\mycolwidth]{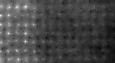}&
    \includegraphics[width=\mycolwidth]{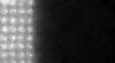}&
    \includegraphics[width=\mycolwidth]{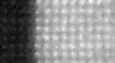}\\
    \CLS{}&
    \includegraphics[width=\mycolwidth]{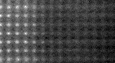}&
    \includegraphics[width=\mycolwidth]{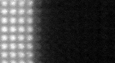}&
    \includegraphics[width=\mycolwidth]{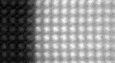}\\
    ITKrMM&
    \includegraphics[width=\mycolwidth]{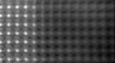}&
    \includegraphics[width=\mycolwidth]{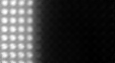}&
    \includegraphics[width=\mycolwidth]{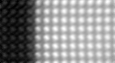}\\
    wKSVD&
    \includegraphics[width=\mycolwidth]{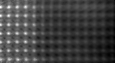}&
    \includegraphics[width=\mycolwidth]{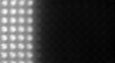}&
    \includegraphics[width=\mycolwidth]{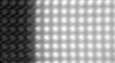}\\
    BPFA&
    \includegraphics[width=\mycolwidth]{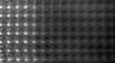}&
    \includegraphics[width=\mycolwidth]{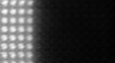}&
    \includegraphics[width=\mycolwidth]{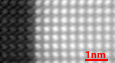}\\
\end{tabular}

%% file: img/spectra_template_real.tex
\setlength{\myfigurewidth}{0.95\columnwidth}
%
\def\fileplot{img/4.real_im/spectra/real_samp.dat}
\subfigure[Sampled pixel (blue pixel)]{
    \begin{tikzpicture}[]
        
        \begin{axis}[
            xlabel={Energy loss (eV)},
            ylabel={Amplitude},
            legend entries={Reference, NN, 3S, \CLS{}, ITKrMM, wKSVD, BPFA},%
            width=\myfigurewidth,
            height=0.5\myfigurewidth,
            ymin=0, ymax=2200,
            %
            legend style={
                draw=none, 
                at={(0.5,1.03)}, 
                anchor=south, 
                legend columns=3, 
                /tikz/every even column/.append style={column sep=0.5cm}
            },
            line width=1pt,
            ]
            \addplot+ [] table [x=eV, y=GT]{\fileplot};
            \addplot+ [] table [x=eV, y=NN]{\fileplot};
            \addplot+ [] table [x=eV, y=SSS]{\fileplot};
            \addplot+ [] table [x=eV, y=FS3D]{\fileplot};
            \addplot+ [] table [x=eV, y=ITKrMM]{\fileplot};
            \addplot+ [] table [x=eV, y=wKSVD]{\fileplot};
            \addplot+ [] table [x=eV, y=BPFA]{\fileplot}; 
            
            \addplot+ [name path=llim, blue, opacity=0.2] coordinates {(829, 0) (829, 2200)};
            \addplot+ [name path=rlim, blue, opacity=0.2] coordinates {(839, 0) (839, 2200)};
            \addplot+ [blue, opacity=0.2] fill between[of=llim and rlim];
        \end{axis}
        
    \end{tikzpicture}
    \label{subfig:real_spectra_samp}
}\\
%
%
\subfigure[Sampled pixel (blue pixel) - zoom]{
    \begin{tikzpicture}[]
        
        \begin{axis}[
            xlabel={Energy loss (eV)},
            ylabel={Amplitude},
            width=\myfigurewidth,
            height=0.6\myfigurewidth,
            xmin=829,
            xmax=839,
            line width=0.5pt,
            ]
            \addplot+ [] table [x=eV, y=GT]{\fileplot};
            \addplot+ [] table [x=eV, y=NN]{\fileplot};
            \addplot+ [] table [x=eV, y=SSS]{\fileplot};
            \addplot+ [] table [x=eV, y=FS3D]{\fileplot};
            \addplot+ [] table [x=eV, y=ITKrMM]{\fileplot};
            \addplot+ [] table [x=eV, y=wKSVD]{\fileplot};
            \addplot+ [] table [x=eV, y=BPFA]{\fileplot}; 
        \end{axis}
    \end{tikzpicture}
    \label{subfig:real_spectra_samp_zoom}
}\\
%
%
\def\fileplot{img/4.real_im/spectra/real_non_samp.dat}
\subfigure[Non-sampled pixel (red pixel)]{
    \begin{tikzpicture}[]
        \begin{axis}[
            xlabel={Energy loss (eV)},
            ylabel={Amplitude},
            width=\myfigurewidth,
            line width=1pt,
            legend entries={Reference, NN, 3S, \CLS{}, ITKrMM, wKSVD, BPFA},
            legend style={
                draw=none, 
                at={(0.5,1.03)}, 
                anchor=south, 
                legend columns=3, 
                /tikz/every even column/.append style={column sep=0.5cm}
            },
            height=0.5\myfigurewidth,
            ymin=0, ymax=2200,
            ]
            \addplot+ [] table [x=eV, y=GT]{\fileplot};
            \addplot+ [] table [x=eV, y=NN]{\fileplot};
            \addplot+ [] table [x=eV, y=SSS]{\fileplot};
            \addplot+ [] table [x=eV, y=FS3D]{\fileplot};
            \addplot+ [] table [x=eV, y=ITKrMM]{\fileplot};
            \addplot+ [] table [x=eV, y=wKSVD]{\fileplot};
            \addplot+ [] table [x=eV, y=BPFA]{\fileplot}; 
            
            \addplot+ [name path=llim, blue, opacity=0.2] coordinates {(829, 0) (829, 2200)};
            \addplot+ [name path=rlim, blue, opacity=0.2] coordinates {(839, 0) (839, 2200)};
            \addplot+ [blue, opacity=0.2] fill between[of=llim and rlim];
        \end{axis}
    \end{tikzpicture}
    \label{subfig:real_spectra_non_samp}
}\\ 
%
%
\subfigure[Non-sampled pixel (red pixel) - zoom]{
    \begin{tikzpicture}[]
        
        \begin{axis}[
            xlabel={Energy loss (eV)},
            ylabel={Amplitude},
            width=\myfigurewidth,
            height=0.6\myfigurewidth,
            xmin=829,
            xmax=839,
            line width=0.5pt,
            ]
            \addplot+ [] table [x=eV, y=GT]{\fileplot};
            \addplot+ [] table [x=eV, y=NN]{\fileplot};
            \addplot+ [] table [x=eV, y=SSS]{\fileplot};
            \addplot+ [] table [x=eV, y=FS3D]{\fileplot};
            \addplot+ [] table [x=eV, y=ITKrMM]{\fileplot};
            \addplot+ [] table [x=eV, y=wKSVD]{\fileplot};
            \addplot+ [] table [x=eV, y=BPFA]{\fileplot}; 
        \end{axis}
    \end{tikzpicture}
    \label{subfig:real_spectra_non_samp_zoom}
}\\

%% file: img/CLS_init.tex
\newcommand{\localtable}{img/5.FS_init/CLS_initialization.dat}
\begin{tikzpicture}
    \begin{axis}[
        xlabel={Iteration number},
        ylabel={SNR},  
        legend entries={Random init., \CLS{} init.},%
        width=\columnwidth,
        height=0.6\columnwidth,
        %
        legend style={draw=none, legend pos=south east},
        ]
        \addplot+ [Dark2-3-1, line width=1pt] table [x=x, y=random_mean]{\localtable};
        \addplot+ [Dark2-3-2, line width=1pt] table [x=x, y=CLS_mean]{\localtable};   
        
        \addplot+ [name path=A, draw=none] table [x=x, y=random_std1] {\localtable};
        \addplot+ [name path=B, draw=none] table [x=x, y=random_std2] {\localtable};
        \addplot+ [name path=C, draw=none] table [x=x, y=CLS_std1] {\localtable};
        \addplot+ [name path=D, draw=none] table [x=x, y=CLS_std2] {\localtable};
        
        \addplot[Dark2-3-1, opacity=0.3] fill between[of=A and B];
        \addplot[Dark2-3-2, opacity=0.3] fill between[of=C and D];
    \end{axis}

\end{tikzpicture}

%% file: sections/A-FS-implementation.tex

\section{\CLS{} implementation}\label{appendix:FS}

This appendix provides some details regarding the implementation of \CLS{} introduced in Section~\ref{subsec:FS-method}. Let $\X\in\mathbb{R}^{\B\times\P}$ denote the unknown spectrum-image to be reconstructed  where \B\ is the number of bands and \P\ is the number of pixels. The subsampling acquisition process writes
\begin{equation}
    \Y = \X\Phi + \E 
\end{equation}
where $\Y\in\mathbb{R}^{\B\times\N}$ is the observation matrix, \N\ is the number of sampled pixels, $\Phi\in\mathbb{R}^{\P\times\N}$ is the sampling operator and \E\ is a residual term associated with error modeling and measurement noise. The elements of \E\ are assumed to be independent and identically distributed according to a zero-mean Gaussian noise. Note that the sampling operator $\Phi$ is the concatenation of \N\ columns extracted from the identity $\P \times \P$-identity matrix.

Reconstructing the full spectrum-image \X\ from \Y\ can be formulated as the following optimization problem
\begin{equation}
    \hat{\X} = \arg\min_{\X}\underbrace{\frac{1}{2}||\Y - \X\Phi||_\mathrm{F}^2}_{f(\X)} + \underbrace{\lambda ||\X\Psi||_{2, 1}}_{g(\X)}
    \label{eq:op}
\end{equation}
where $\Psi$ is the band-by-band DCT transform operator. This optimization problem can be easily solved with iterative algorithms such as the fast iterative shrinkage-thresholding algorithm (FISTA)~\cite{beck2009fast}.  This algorithm splits the function to be minimized into two terms
\begin{itemize}
    \item $f$ which is a smooth convex function of the type $\mathcal{C}^{1, 1}$, i.e., continuously differentiable with Lipschitz continuous gradient $L(f )$,
    \item $g$ which is a continuous convex and possibly nonsmooth function.
\end{itemize}
Based on this decomposition, the main steps of the algorithm are described in Algo.~\ref{algo:FISTA}, which requires to evaluate three elements, namely, the gradient $\nabla f$, an upper bound of the Lipschitz constant $L>L(f)$ and the proximal operator \me{$\mathrm{prox}_{g/L}(\X)=\arg\min_\U \{ g(\U)/L + \frac{1}{2}||\U - \X||^2 \}$}. The gradient function $\nabla f$ can be easily derived as
\begin{equation}
    \nabla f (\X) = (\X\Phi - \Y)\Phi^T.
\end{equation}
The $\nabla f$ function Lipschitz constant can be computed following
\begin{equation}
    ||\nabla f(\X_1) - \nabla f(\X_2)|| = ||\X_1-\X_2||\cdot ||\Phi\Phi^T||
\end{equation}
with $L(f)=||\Phi\Phi^T|| = 1$. 
Last, the proximal operator is column-separable and is computed for each column $\X_j$ of $\X$ (with $j=1,\ \dots,\ \P$) \cite{jenatton2011proximal}
\me{
\begin{equation}
    [\mathrm{prox}_{g/L}(\X)]_j = 
    \begin{cases}
        0&\text{if}\ ||\X_j||_2 < \frac{\lambda}{L}\\
        \left(1-\frac{\lambda/L}{||\X_j||_2}\right)\X_j&\text{otherwise}\\
    \end{cases}\label{eq:prox-op}
\end{equation}
}

As mentioned in Section~\ref{subsec:FS-method}, the reconstruction method can be applied to the full measurement matrix $\Y$ or to its representation $\tilde{\Y}$ in a lower-dimensional subspace identified by PCA. The threshold $T$ to be set can be chosen as described in the following section.

\begin{algorithm}\label{algo:FISTA}
    \caption{FISTA with constant step size \cite{beck2009fast}}
    \DontPrintSemicolon
    \KwSty{Input :}\ {$L> L_{f}$ an upper bound of $L_{f}$ \corr{and the observation \Y}}\;
    \KwSty{Initialisation :}\ Set $\corr{\mathbf{Z}^{(1)} = \mathbf{X}^{(0)}} \in \mathbb{R}^p$, $\theta^{(1)}=1$, $i=1$
    \While{stopping rule not satisfied}{
        %
        %
        $\corr{\mathbf{X}^{(i)}} = \mathrm{prox}_{g/L}\left( \corr{\mathbf{Z}^{(i)}-\frac{1}{L}(\mathbf{Z}^{(i)}\Phi - \Y)\Phi^T} \right)$\label{LigneAlgo1} \label{algostep:gradient}\;
        $\theta^{(i+1)} = \frac{1}{2} \left(1+\sqrt{1+4(\theta^{(i)})^2}\right)$\label{algostep:theta-update}\;
        $\corr{\mathbf{Z}^{(i+1)}} = \corr{\mathbf{X}^{(i)}} + \left( \frac{\theta^{(i)}-1}{\theta^{(i+1)}} \right) \left(\corr{\mathbf{X}^{(i)}-\mathbf{X}^{(i-1)}} \right)$\label{algostep:y-update}\;
        $i \leftarrow i+1$\label{algostep:increment}
    }
\end{algorithm} 

%% file: sections/B-Synthetic-details.tex

\section{Generation of the synthetic and semi-real spectrum-images}\label{appendix:synth-details}

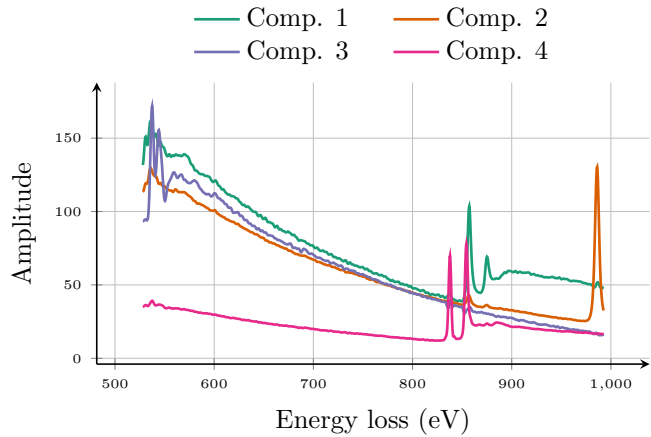
\begin{figure}[htbp]
    \centering
    \extcode{%
        \begin{tikzpicture}
            \begin{axis}[
                xlabel={Energy loss (eV)},
                ylabel={Amplitude},
                legend entries={Comp. 1, Comp. 2, Comp. 3, Comp. 4},
                width=\columnwidth,
                height=0.6\columnwidth,
                %
                legend style={
                    draw=none, 
                    at={(0.5,1.03)}, 
                    anchor=south, 
                    legend columns=2, 
                    /tikz/every even column/.append style={column sep=0.5cm}
                },
                ]
                
                \addplot+ [no markers] file {img/B.Synth/Synth_creation/spectrum_0.dat};
                \addplot+ [no markers] file {img/B.Synth/Synth_creation/spectrum_1.dat};
                \addplot+ [no markers] file {img/B.Synth/Synth_creation/spectrum_2.dat};
                \addplot+ [no markers] file {img/B.Synth/Synth_creation/spectrum_3.dat};
            \end{axis}
        \end{tikzpicture}
    }{%
        \includegraphics[scale=1]{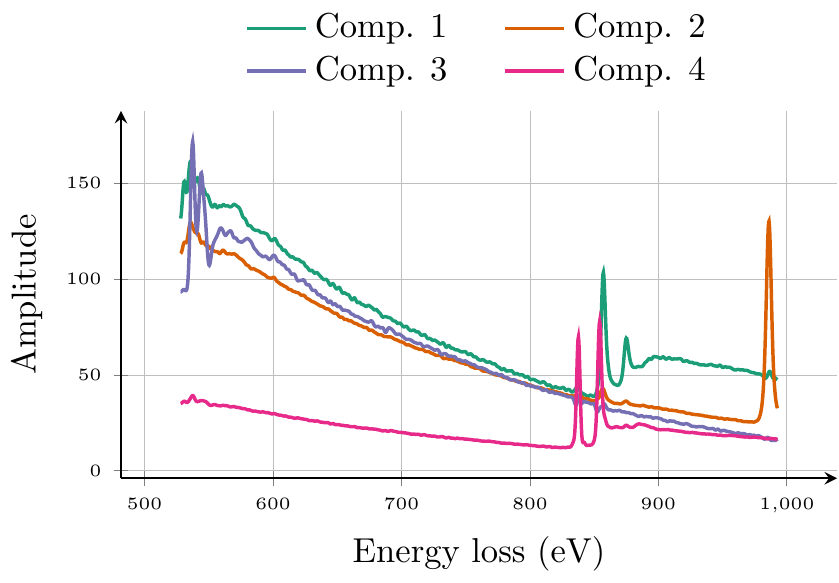}
    }
    \caption{Spectral components used for generating the synthetic image $\mathsf{S}$.}
    \label{fig:spectra-synth}
\end{figure}

\begin{figure}[htbp]
    \centering
    \setlength\mycolwidth{0.45\columnwidth}
    \subfigure[Comp. 1]{\includegraphics[width=\mycolwidth]{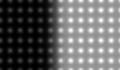}}
    \subfigure[Comp. 2]{\includegraphics[width=\mycolwidth]{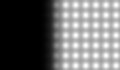}}
    \subfigure[Comp. 3]{\includegraphics[width=\mycolwidth]{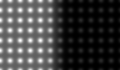}}
    \subfigure[Comp. 4]{\includegraphics[width=\mycolwidth]{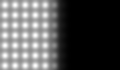}}
    \caption{Proportion maps associated with the spectral components represented in Fig.~\ref{fig:spectra-synth} used for generating the synthetic image $\mathsf{S}$. }
    \label{fig:maps-synth}
\end{figure}

\paragraph{Generating the synthetic spectrum-image $\mathsf{S}$} 
To generate the synthetic spectrum-image $\mathsf{S}$, we propose to follow the strategy adopted in \cite{monier2018tci}. It consists in linearly mixing $\Nc=4$ components according to realistic proportion maps. The spectral components were obtained by smoothing $\Nc$ signatures identified by an independent component analysis conducted on the real-spectrum image $\mathsf{R}_2$. They are gathered in the $B\times \Nc$ matrix $\M$ and depicted  in Fig.~\ref{fig:spectra-synth}. The proportion maps were generated synthetically by superimposing periodic structures onto a smoothly varying background similar to the content of the real spectrum-image $\mathsf{R}_2$. These maps, gathered in the $\Nc \times P$ matrix $\A$, are represented in Fig.~\ref{fig:maps-synth}. Finally, the synthetic image $\mathsf{S}$ can be generated as  $\M\A$.

\begin{figure}[htbp]
    \centering
    \extcode{%
        \begin{tikzpicture}
            \begin{axis}[
                xmode=log,
                ymode=log,
                xlabel={Principal component index},
                ylabel={$||r||_2^*$},
                legend entries={$\mathsf{R}_1$, $\mathsf{R}_2$},  
                width=\columnwidth,
                height=0.6\columnwidth,
                %
                legend style={draw=none, legend pos=north east},
                ]
                \addplot+ [no markers] file {img/1.data_PCA_th/Curves_whiteness_HR1.dat};
                \addplot+ [no markers] file {img/1.data_PCA_th/Curves_whiteness_HR2.dat};
            \end{axis}
        \end{tikzpicture}
    }{%
        \includegraphics[scale=1]{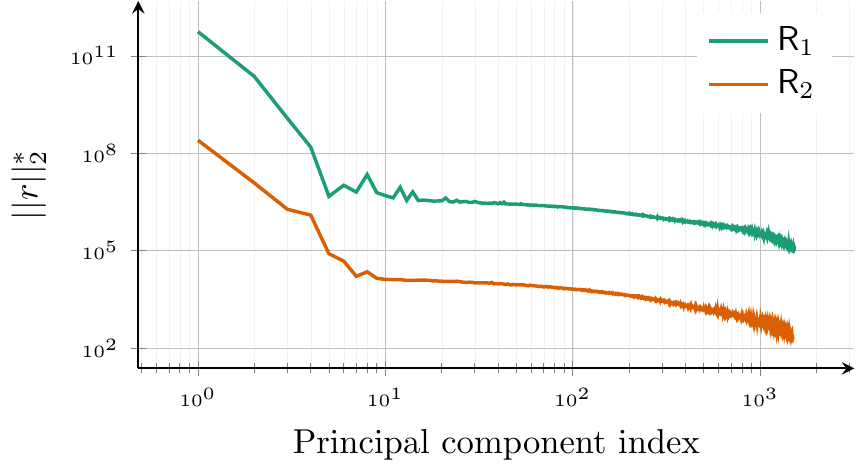}
    }
    \caption{Whiteness criterion $||r||_2^*$ as a function of the principal component index for the real image $\mathsf{R}_1$ and $\mathsf{R}_2$. The first powerful principal components exhibit more spatial content than the last ones. \corr{The PCA threshold is chosen as the maximal index sufficient to reach a stationary curve behavior.}}
    \label{fig:whiteness-metric}
\end{figure}
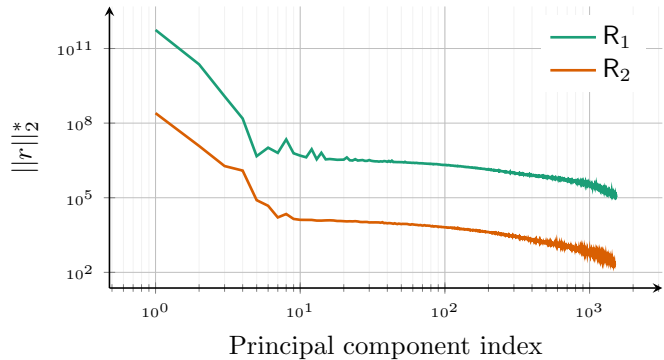

\paragraph{Generating the semi-real spectrum-images $\mathsf{R}_1^*$ and $\mathsf{R}_2^*$} 
The reference noise-free spectrum-images $\bar{\mathsf{R}}_1^*$ and  $\bar{\mathsf{R}}_2^*$ used in Section~\ref{subsec:synthetic-results} are generated by keeping the first $T$ principal components identified by PCA conducted on $\mathsf{R}_1$ and $\mathsf{R}_2$, respectively. Generally, determining the PCA threshold $T$ is not a trivial. In this work, we propose to choose $T$ such as the $\B-T$ remaining components do not contain spatial information. To quantify the presence or absence of spatial information in a principal components, we monitor their whiteness based on the metrics proposed in \cite[Chap. 3]{riot2018residual} as
\begin{equation}
     ||r||_2^* = \sqrt{\sum_{\tau\neq 0} r(\tau)^2}
 \end{equation} 
where $r(\tau)$ is the 2D-autocorrelation function. The higher value, the more information contained by the image. To illustrate, this criterion is represented in Figure~\ref{fig:whiteness-metric} as a function of the index of the principal components for the two spectrum-images $\mathsf{R}_1$ and $\mathsf{R}_2$. The PCA threshold $T$ is chosen as the maximal index sufficient to get to a stationary curve behavior. These values are reported in Table~\ref{table:samples}. The semi-real spectrum-images $\mathsf{R}_1^*$ and $\mathsf{R}_2^*$ are finally obtained by corrupting the reference noise-free spectrum-images $\bar{\mathsf{R}}_1$ and  $\bar{\mathsf{R}}_2$ with additive white Gaussian noises whose variances have been adjusted to reach SNR values in agreement with those of the corresponding real spectrum-images $\mathsf{R}_1$ and $\mathsf{R}_2$, respectively.

